\documentclass[usenatbib]{mn2e}
\usepackage[dvips]{graphicx}
\usepackage{amssymb,amsmath}
\usepackage[titletoc]{appendix}
\usepackage{times}
\usepackage{fleqn,subfigure,lscape,epsfig,graphics}
\title[The \textit{WISE} properties of complete samples of radio-loud AGN]{The \textit{WISE} properties of complete samples of radio-loud AGN}
\author[G\"urkan et al.]
{G. G\"urkan $^{1}$\thanks{E-mail:g.gurkan-uygun@herts.ac.uk},
M.J. Hardcastle$^{1}$ and
M.J. Jarvis$^{2,3}$
\\
$^{1}$ School of Physics, Astronomy and Mathematics, University of Hertfordshire, College Lane, Hatfield AL10 9AB\\
$^{2}$ Astrophysics, Department of Physics, Keble Road, Oxford, OX1 3RH\\
$^{3}$ Physics Department, University of the Western Cape, Private Bag X17, Bellville 7535, South Africa}
\begin{document}
% bibliography and bibfile journal definitions (taken from aa.cls)

% Astronomy and Astrophysics
\def\aj{AJ}					% Astronomical Journal
\def\araa{ARA\&A}				% Annual Reviews of Astronomy and Astrophysics
\def\nar{NewAR}                                 % New Astronomy Reviews
\def\apj{ApJ}					% Astrophysical Journal
\def\apjl{ApJL}					% Astrophysical Journal, Letters
\def\apjs{ApJS}					% Astrophysical Journal, Supplement Series
\def\apss{Astrophysics and Space Science}
\def\capsp{Comments on Astrophysics and Space Physics}
\def\aap{A\&A}					% Astronomy and Astrophysics
\def\aapr{A\&A~Rev.}				% Astronomy and Astrophysics Reviews
\def\aaps{A\&AS}				% Astronomy and Astrophysics, Supplement
\def\azh{AZh}					% Astronomicheskii Zhurnal
\def\baas{BAAS}					% Bulletin of the AAS
\def\jrasc{JRASC}				% Journal of the RAS of Canada
\def\memras{MmRAS}				% Memoirs of the RAS
\def\mnras{MNRAS}					% Monthly Notices of the Royal Astronomical Society
\def\pasp{PASP}					% Publications of the ASP
\def\pasj{PASJ}					% Publications of the ASJ
\def\qjras{QJRAS}				% Quarterly Journal of the RAS
\def\skytel{S\&T}				% Sky and Telescope
\def\solphys{Sol.~Phys.}			% Solar Physics
\def\sovast{Soviet~Ast.}			% Soviet Astronomy
\def\ssr{Space~Sci.~Rev.}			% Space Science Reviews
\def\zap{ZAp}					% Zeitschrift fuer Astrophysik
\def\na{New Astronomy}				% New Astronomy
\def\iaucirc{IAU~Circ.}				% IAU Cirulars
\def\aplett{Astrophys.~Lett.}			% Astrophysics Letters
\def\apspr{Astrophys.~Space~Phys.~Res.}		% Astrophysics Space Physics Research
\def\bain{Bull.~Astron.~Inst.~Netherlands}	% Bulletin Astronomical Institute of the Netherlands
\def\memsai{Mem.~Soc.~Astron.~Italiana}		% Mem. Societa Astronomica Italiana

% Optics
\def\ao{Appl.~Opt.}				% Applied Optics

% General physics
\def\pra{Phys.~Rev.~A}				% Physical Review A: General Physics
\def\prb{Phys.~Rev.~B}				% Physical Review B: Solid State
\def\prc{Phys.~Rev.~C}				% Physical Review C
\def\prd{Phys.~Rev.~D}				% Physical Review D
\def\pre{Phys.~Rev.~E}				% Physical Review E
\def\prl{Phys.~Rev.~Lett.}			% Physical Review Letters
\def\nat{Nature}				% Nature
\def\fcp{Fund.~Cosmic~Phys.}			% Fundamental Cosmic Physics
\def\gca{Geochim.~Cosmochim.~Acta}		% Geochimica Cosmochimica Acta
\def\grl{Geophys.~Res.~Lett.}			% Geophysics Research Letters
\def\jcp{J.~Chem.~Phys.}			% Journal of Chemical Physics
\def\jgr{J.~Geophys.~Res.}			% Journal of Geophysics Research
\def\jqsrt{J.~Quant.~Spec.~Radiat.~Transf.}	% Journal of Quantitiative Spectroscopy and Radiative Trasfer
\def\nphysa{Nucl.~Phys.~A}			% Nuclear Physics A
\def\physrep{Phys.~Rep.}			% Physics Reports
\def\physscr{Phys.~Scr}				% Physica Scripta
\def\planss{Planet.~Space~Sci.}			% Planetary Space Science
\def\procspie{Proc.~SPIE}			% Proceedings of the SPIE
\def\rpp{Rep.~Prog.~Phys.}			% Rep. Prog. Phys.
\let\astap=\aap
\let\apjlett=\apjl
\let\apjsupp=\apjs
\let\applopt=\ao
\let\prep=\physrep

% end of file

\date{Accepted ...... Received ...... ; in original form......   }

\pagerange{\pageref{firstpage}--\pageref{lastpage}} \pubyear{2011}
\maketitle
\label{firstpage}
\begin{abstract}

We present an analysis of four complete samples of radio-loud AGN (3CRR, 2Jy, 6CE and 7CE) using near- and mid-IR data taken by the $\textit{Wide-Field Infrared Survey Explorer}$ ($\textit{WISE}$). The combined sample consists of 79 quasars and 273 radio galaxies, and covers a redshift range $0.003<z<3.395$. The dichotomy in the mid-IR properties of low- and high-excitation radio galaxies (LERGs - HERGs) is analysed for the first time using large complete samples. Our results demonstrate that a division in the accretion modes of LERGs and HERGs clearly stands out in the mid-IR$-$radio plane ($L_{22 \mu m}$=5$\times$10$^{43}$ erg s$^{-1}$). This means that $\textit{WISE}$ data can be effectively used to diagnose accretion modes in radio-loud AGN. The mid-IR properties of all objects were analysed to test the unification between quasars and radio galaxies, consistent with earlier work and we argue that smooth torus models best reproduce the observation. Quasars are found to have higher mid-IR luminosities than radio galaxies. We also studied all the sources in the near-IR to gain insights into evolution of AGN host galaxies. A relation found between the near-IR luminosity and redshift, well-known in the near-IR, is apparent in the two near-IR $\textit{WISE}$ bands, supporting the idea that radio sources are hosted by massive elliptical galaxies that formed their stars at high redshifts and evolved passively thereafter. Evaluation of the positions of the sample objects in $\textit{WISE}$ colour-colour diagrams shows that widely used $\textit{WISE}$ colour cuts are not completely reliable in selecting AGN.

\end{abstract}


\section{INTRODUCTION}
The nature of the energy source in active galactic nuclei (AGN) has been a matter of debate for a long while. In commonly accepted models, the energy is generated by accretion of cold material onto black holes in the centre of active galaxies \citep{1995ref64,1998ref65,2000ref66}. The accretion produces photoionizing ultra-violet (UV) radiation, and gives rise to X-ray emission via Compton scattering. Hot, high-velocity gas clouds located within $\sim$ 1pc of the obscuring torus produce the broad emission lines that can be observed in their spectra while narrow emission lines are produced by the lower velocity gas clouds situated further away at $\sim$ 10-100 pc \citep{1984ref18,1995ref16}. Different classes of AGN, however, present different observational features in optical, X-ray and radio bands. An orientation effect is the main ingredient of both optical and radio unification schemes: in the optical, radio-loud AGN are classified according to whether they have broad emission lines in their spectra, which can be obscured by dust and gas (torus) at certain angles \citep[e.g.][]{1985ref39}. In this case, the obscuring structure is expected to re-radiate strongly in the mid-infrared \citep{2001ref21,2004ref19,2006ref20}.

The simple unification scheme of quasars and radio galaxies developed by \cite{1989ref17} is based on orientation dependent effects. According to this model, radio galaxies and quasars are the same objects seen at different angles. If a source is viewed within a cone of half-angle approximately 45$^{\circ}$ it is called a quasar or broad-line radio galaxy (BLRG); if a source is viewed edge-on, where the nucleus is blocked by the dusty torus, the source is then identified as a radio galaxy (narrow-line radio galaxy - NLRG). On the other hand, \cite{1979ref22} pointed out the existence of a population of radio galaxies that do not present the strong emission lines conventionally seen in powerful AGN (high-excitation radio galaxies - HERGs). This class of objects, called low-excitation radio galaxies (LERGs), are predominantly found to be at low-radio luminosities. This population poses a problem for simple unification, falling outside the generic view of radio-loud AGN. They do not exhibit any expected feature of unified AGN; radiatively efficient accretion disk, X-ray emitting corona \citep{2006ref28,2006ref25} and obscuring torus in the mid-IR \citep{2004ref19,2006ref20}. 

\cite{2007ref27} suggested that a different source of fuel for the accretion process may be responsible for this difference. In this picture, HERGs (quasars, BLRGs and NLRGs) accrete in `cold mode'; in which accretion of cold matter onto a super-massive black hole via an optically thick geometrically thin accretion disk produce radiation efficiently, whereas LERGs (alternatively called weak-line radio galaxies, or WLRGs) are thought to be fuelled by the accretion of the hot gas in haloes of their host galaxies through advection dominated flows \citep{1995ref34,2006ref67,2007ref27,2012ref68} and release the accretion energy in the form of jets or winds \citep{2002ref26,2007ref30} [see also \cite{2011ref31,2011ref32}]. The relation between accretion rate and jet power was studied by \cite{2006ref33} and Bondi accretion rates have been evaluated for LERGs and HERGs by \cite{2007ref27}. They showed that the majority of LERGs have jet powers comparable to the available Bondi power while many NLRGs have jet powers higher than Bondi accretion level, emphasising the possibility that LERGs are powered by accretion from the hot phase. Recently a similar conclusion was reached by \cite{2013ref90}. It is worth noting that studies of AGN environments support the idea of a different origin of the accreting gas; LERGs occupy gas-rich environments and redder, lower star-formation galaxies in comparison to HERGs at comparable redshifts \citep[e.g][]{2004ref35,2007ref62,2008ref42,2012ref29,2013ref82}. However, another hypothesis to explain the diversity in accretion-modes of AGN is that there is a limiting value of the Eddington-scaled accretion rate above which radiatively efficient accretion takes place \citep[e.g.][]{1995ref34}. It has recently been shown that LERGs and HERGs are plausibly separated at a critical value of the Eddington-scaled accretion rate \citep[e.g.][Mingo et al. 2013]{2012ref29,2013ref90}. This model does not require a one-to-one correspondence between fuel source and accretion mode, but, since sources fuelled by accretion from the hot phase will tend to contain massive black holes being fuelled at a low rate, it retains many of the predictions from the \cite{2007ref27} model. Further discussion of the nature of LERGs and HERGs can be found in a recent review by \cite{2012ref120}.

Determining the radiative power of AGN, particularly obscured AGN, is observationally difficult. To date, many hidden AGN have been discovered using hard X-rays but this technique only works for Compton-thin AGN \citep{2000ref40,2001ref41}. Spectropolarimetry is another way of confirming a hidden nucleus \citep{1983ref38,1985ref39,1997ref74}. However, it has limitations because in the case of lack of scattering material the results would yield false negatives.
Mid-IR data is important in searching for radiatively efficient accretion in radio galaxies and testing the unification hypothesis, because in the presence of the putative dusty torus, optical-UV emission from quasar nucleus is intercepted by the torus and re-radiated in the mid-IR, which is only mildly sensitive to orientation \citep{2004ref19,2005ref52,2009ref51,2009ref10,2011ref114}. 

Many previous authors have used the mid-IR properties of radio sources to test the unification hypothesis. \cite{1992ref69,1994ref56} used the Infrared Astronomical Satellite ($\textit{IRAS}$) measurements of 3CR radio galaxies and quasars at z$>$0.3, and found that quasars were $\approx$ 4 times brighter than radio galaxies in their mid-IR and far-infrared (far-IR) fluxes (at 25-, 60- and 100-$\mu$m). They concluded that either mid-IR emission from radio-loud AGN is not isotropic or quasars are intrinsically different sources of mid-IR than radio galaxies with the same radio power, which would imply that the simple unification scheme needs to be revised. Similar results were also obtained in various subsequent studies using $\textit{Spitzer}$ \citep[e.g.][]{2007ref43,2008ref86} and different reasons were suggested for the excess magnitudes seen in quasars in the mid-IR such as non-thermal emission and dust absorption. On the other hand, some other studies provided opposite results, suggesting that quasars and radio galaxies show identical mid-IR magnitudes \citep[e.g.][]{2001ref21,2004ref70,2005ref71,2005ref87,2005ref52,2009ref10}.

Mid-IR studies of radio sources hitherto have been based on single sources or small or incomplete samples. In addition to this, due to the lack of complete mid-IR observations of radio samples, many investigations were carried out using incoherent data sets where the mid-IR observations were taken for a variety of purposes. One of the motivations of the current paper is to remove this deficit in radio-loud AGN research, providing a complete mid-IR study of the complete and relatively large sample. Here we present, for the first time, the mid-IR properties of four complete samples, namely 3CRR, 2Jy, 6CE and 7CE, which overall cover a wide redshift range ($0.0029<z<3.93$). For the analysis, the mid-IR data taken by $\textit{Wide-Field Infrared Survey Explorer}$ ($\textit{WISE}$) \citep{2010ref44} have been used. The objective of this work is to evaluate the utility of mid-IR data to investigate radiatively efficient/inefficient radio loud-AGN and the nature of HERGs and LERGs with $\textit{WISE}$ data. We also examine the mid-IR properties of these complete samples by addressing the quasar-radio galaxy unification and the near-IR properties by comparing our results with previous studies.

This paper is organised as follows: a description of the samples, the method of determination of $\textit{WISE}$ fluxes and luminosities and tests of the consistency between $\textit{WISE}$ and $\textit{Spitzer}$ data are given in Section 2. Our key results are given in Section 3: $\textit{WISE}$ colour-colour diagrams of the sources are presented discussing the available colour cuts in the literature. An investigation of the old stellar population in our sample is presented and the unification of quasars and NLRGs is discussed. The dichotomy for LERGs and HERGs is re-evaluated in the light of our results. Finally, Section 4 presents a summary of our results and conclusions drawn.

The cosmological parameters used throughout the paper are as follows: $\Omega_{m}$=0.3, $\Omega_{\Lambda}$=0.7 and $H_{0}$=70 km s$^{-1}$ Mpc$^{-1}$. 
%%%%%%%%%%%%%%%%%%%%%%%%%%%%%%%%%%%%%%%%%%%%%%%%%%%%%%%%%%%%%%%%%%%%%%%%%%%%%%%%%%%%%%%

\section{DATA and ANALYSIS}
 \subsection{Samples}
\begin{figure}
\begin{center}
\scalebox{0.89}{
\begin{tabular}{cc}
\centerline{\hspace{-0.9em}\includegraphics[width=10cm,height=10cm,angle=0,keepaspectratio]{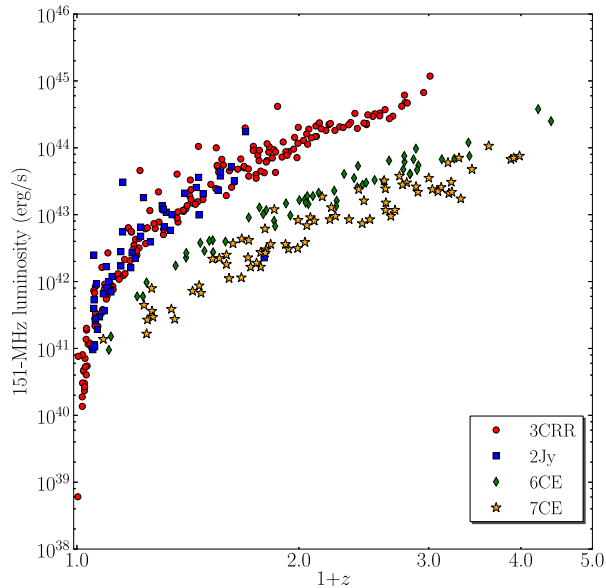}}\\
\end{tabular}}
\caption[.]{The 151-MHz radio luminosity distribution of the samples against redshift described in section 2.1. The 3CRR sample objects are plotted as red circles, the 2Jy sample as blue squares, the 6CE sample as orange stars and the 7CE sample as green diamonds. \label{sources}}
\end{center}
\end{figure}

The 3CRR, 2Jy, 6CE and 7CE samples were chosen for our analysis. These complete samples are designed to include all radio sources brighter than the specified flux density limit in a particular area of sky at the selected frequency. Most of the samples have complete redshift measurements and emission-line classifications. Figure \ref{sources} shows the low-frequency radio luminosity and redshift distributions of the objects in the four samples which are described in more detail in the following subsections.

\subsubsection{The 3CRR Sample:}We used the revised sub-sample of the 3CR catalogue of radio sources \citep{1962ref1}, which have flux densities greater than 10.9 Jy at 178 MHz \citep{1983ref2}. There are 172 sources with $0.0029<z<2.012$ including 37 LERGs, 82 NLRGs, 10 BLRGs and 43 quasars. 3C231 was excluded from the analysis as its radio emission is due to a starburst. 

\subsubsection{The 2Jy Sample:}Another complete radio sample chosen for our analysis is a sub-sample of the 2Jy objects which has homogeneous spectroscopic observations \citep{1985ref3,1993ref4}. The complete sample was generated selecting radio objects with flux densities above 2Jy at 2.7 GHz. Our sub-sample consists of 48 steep-spectrum sources (objects dominated by emission from the beamed relativistic jet and core components are excluded so that we minimise the contamination from non-thermal emission) with $0.05<z<0.7$, which have mid-far infrared (MFIR) imaging and spectra taken by $\textit{Spitzer}$ \citep{2008ref11,2009ref10,2012ref9} as well as 98 percent complete $\textit{Chandra/XMM}$ X-ray imaging (Mingo et.al 2013 (in prep.)). This sample has 10 LERGs, 20 NLRGs, 13 BLRGs and 5 quasars.

\subsubsection{The 6CE Sample:}The 6CE sub-sample is drawn from the 6C survey \citep{1985ref7} and was designed to select objects fainter than 3C objects, in order to investigate the cosmic evolution of radio galaxies \citep{1985ref8,1997ref6}. The flux density limit for the sample is 2$\leq$ S$_{151}$$\leq$3.93 Jy. It has virtually complete spectroscopic redshift measurements as well as infrared imaging \citep{1997ref6,2001ref5}. The 6CE sample has 58 sources including 19 LERGs, 28 HERGs and 9 quasars, whose redshift ranges between 0.105 and 3.395.

\subsubsection{The 7CE Sample:}The 7CE sample, drawn from the 7C survey \citep{1995ref12}, is a complete sample having flux densities greater than 0.5 Jy at 151 MHz \citep{1998ref13,2002ref14}. This sample has 90 percent spectroscopic redshift completeness and 49 radio galaxies have near-infrared imaging \citep{2003ref15}. It comprises 74 radio sources covering redshift from 0.086 to 2.982: 4 LERGs, 46 NLRGs, 2 BLRGs and 22 quasars. 

Radio sources in the 7CE sample are classified only as NLRGs, BLRGs and quasars. In order to distinguish LERGs from NLRGs, equivalent widths of the [OII] and [OIII] emission lines were collected from the literature \citep{1998ref13,2003ref15}. NLRGs were re-classified as LERGs considering the criteria suggested by \cite{1997ref50}: sources with equivalent widths of $<$10\,\AA\, in [OIII] or [OII]/[OIII] ratios of $>$1 or both. Many NLRGs do not have [OIII] detections in their spectra; this is mainly because at higher redshifts the [OIII] is redshifted beyond the visible spectral window. For this reason, we note that there may be more LERGs in the 7CE sample classified as NLRGs at $z>0.8$.

\subsection{$\textit{WISE}$ magnitudes}
The $\textit{WISE}$ mission has observed the whole sky in four mid-IR bands ($W1$ [3.4-$\mu$m], $W2$ [4.6-$\mu$m], $W3$ [12-$\mu$m], $W4$ [22-$\mu$m]) with an angular resolution of 6.1, 6.4, 6.5 and 12 arcsec, respectively. The $\textit{WISE}$ all-sky catalogue was searched for all objects in our samples. This was done by searching the catalogue within 10 arcsec.The expected number of sources within 10 arcsec is around 3.49$\times$10$^{-4}$ so that a source detected within the search region of a given position is very unlikely to be a false association: the total expected number of false detections is 0.12. Our sample sources are away from the Galactic plane, where the density of $\textit{WISE}$ sources is highest. Since powerful radio galaxies tend to be the dominant objects in any group or clusters they inhabit, we do not expect any excess source density due to clustering to affect our results. In the search, we used the optical coordinates of the 6CE \citep{1982ref45,1985ref47,1989ref46,1993ref48} and 7CE samples \citep{1998ref13,2003ref15}, and radio coordinates of the 3CRR\footnote{http://3crr.extragalactic.info/} and 2Jy samples\footnote{http://2jy.extragalactic.info/} (radio core coordinates provide accurate positions of AGN but the 6CE and 7CE samples do not have high-resolution radio observations capable of resolving the radio cores of the objects). In the case of multiple matches per source, these objects were treated separately; $\textit{WISE}$ images of individual sources were obtained and checked against high-resolution radio images to make sure that the right source was detected. Matches for all of the 3CRR (172) and 2Jy (48) objects were found in the catalogue. However, only 47 sources in the 6CE sample out of 58 and 68 in the 7CE sample (74) had detections. Sources rejected from the $\textit{WISE}$ all-sky catalogue for various reasons (such as low flux signal-to-noise ratios, spurious detections of image artifacts, or duplicate entries of source detections) are stored in the $\textit{WISE}$ all-sky reject table. The $\textit{WISE}$ reject table was searched to get magnitudes of non-detected sources. In our case, the sources found in the reject table have either upper limits or low flux signal-to-noise ratio. Among 17 non-detected sources 15 of them had matches or upper limit measurements in the reject table. $\textit{WISE}$ images of the remaining two sources were checked. Because of a very low signal-to-noise (SNR), we were not able to obtain their magnitudes, so these sources (6CE-1143+3703 and 6CE-1148+3638) were excluded from the analysis. Almost all of the sources in these samples had matches with SNR$>$2 in the shorter $\textit{WISE}$ bands (3.4 and 4.6 $\mu$m). Most of the upper-limit measurements (SNR$<$2) were obtained in the 12 and 22-$\mu$m $\textit{WISE}$ bands. Upper limits are indicated as arrows in the plots presented in the rest of the paper. Our detections are summarised in Table \ref{samples}, which also presents the quantity of upper-limit measurements grouped according to source classification. The $\textit{WISE}$ measurements are given in Vega magnitudes so the magnitudes of the sources were converted into Jy using the standard $\textit{WISE}$ zero-points\footnote{The relation used for the conversion can be found at $http://wise2.ipac.caltech.edu/docs/release/allsky/expsup/$}.

\begin{table*}
\begin{tabular}{cccccccccc}
\hline
\hline
Source&$z$&Mag$_{(3.4\mu m)}$&Error$_{(3.4\mu m)}$&Mag$_{(4.6\mu m)}$&Error$_{(4.6\mu m)}$&Mag$_{(12\mu m)}$&Error$_{(12\mu m)}$&Mag$_{(22\mu m)}$&Error$_{(22\mu m)}$\\
name&&(Vega)&(Vega)&(Vega)&(Vega)&(Vega)&(Vega)&(Vega)&(Vega)\\
\hline
\hline
4C12.03&0.156&13.345&0.027&13.158&0.033&11.759&0.221&8.626&0.0\\
3C6.1&0.840&14.667&0.031&13.952&0.037&11.528&0.131&8.848&0.353\\  
3C9&2.012&13.986&0.031&12.969&0.038&10.233&0.075&8.509&0.354\\    
3C13&1.351&15.685&0.047&15.02&0.076&11.463&0.129&8.505&0.219\\   
3C14&1.469&13.816&0.029&12.454&0.027&9.377&0.036&7.136&0.099\\
.&.&.&.&.&.&.&.&.&.\\
.&.&.&.&.&.&.&.&.&.\\
.&.&.&.&.&.&.&.&.&.\\
\hline
\end{tabular}
\caption[]{Full tables of $\textit{WISE}$ magnitudes of the sample are available online from the journal website.} {\label{mags}}
\end{table*}

\begin{table*}
\begin{tabular}{cccccccccc}
\hline
\hline
Source&$z$&F$_{(3.4\mu m)}$&Error$_{(3.4\mu m)}$&F$_{(4.6\mu m)}$&Error$_{(4.6\mu m)}$&F$_{(12\mu m)}$&Error$_{(12\mu m)}$&F$_{(22\mu m)}$&Error$_{(22\mu m)}$\\
name&&(Jy)&(Jy)&(Jy)&(Jy)&(Jy)&(Jy)&(Jy)&(Jy)\\
\hline
\hline
4C12.03&0.156&1.41$\times10^{-3}$&4.08$\times10^{-5}$&9.34$\times10^{-4}$&3.16$\times10^{-5}$&5.74$\times10^{-4}$&1.17$\times10^{-4}$&2.9$\times10^{-3}$&0.0\\

3C6.1&0.840&4.16$\times10^{-4}$&1.34$\times10^{-5}$&4.48$\times10^{-4}$&1.67$\times10^{-5}$&7.10$\times10^{-4}$&8.64$\times10^{-5}$&2.39$\times10^{-3}$&7.78$\times10^{-4}$\\

3C9&2.012&7.80$\times10^{-4}$&2.51$\times10^{-5}$&1.11$\times10^{-3}$&4.22$\times10^{-5}$&2.34$\times10^{-3}$&1.65$\times10^{-4}$&3.2$\times10^{-3}$&1.07$\times10^{-3}$\\

3C13&1.351&1.63$\times10^{-4}$&7.47$\times10^{-6}$&1.67$\times10^{-4}$&1.20$\times10^{-5}$&7.54$\times10^{-4}$&9.03$\times10^{-5}$&3.28$\times10^{-3}$&6.63$\times10^{-4}$\\

3C14&1.469&9.12$\times10^{-4}$&2.79$\times10^{-5}$&1.78$\times10^{-3}$&5.19$\times10^{-5}$&5.19$\times10^{-3}$&1.87$\times10^{-4}$&1.15$\times10^{-2}$&1.07$\times10^{-3}$\\   
.&.&.&.&.&.&.&.&.&.\\
.&.&.&.&.&.&.&.&.&.\\
.&.&.&.&.&.&.&.&.&.\\
\hline
\end{tabular}
\caption[]{Full tables of $\textit{WISE}$ fluxes of the sample are available online from the journal website.} {\label{fluxes}}
\end{table*}

\begin{table*}
\begin{tabular}{ccccccc}
\hline
\hline
Samples&Class&Quantity&3.4-$\mu$m&4.6-$\mu$m&12-$\mu$m&22-$\mu$m\\
&&&upper-limits&upper-limits&upper-limits&upper-limits\\
\hline
\hline
3CRR&Total&172&$-$&$-$&21&41\\
&LERG&37&$-$&$-$&6&12\\
&NLRG&82&$-$&$-$&14&26\\
&BLRG&10&$-$&$-$&$-$&$-$\\
&Quasar&43&$-$&$-$&1&2\\
2Jy&Total&48&$-$&$-$&$-$&3\\
&LERG&10&$-$&$-$&$-$&3\\
&NLRG&20&$-$&$-$&$-$&$-$\\
&BLRG&13&$-$&$-$&$-$&$-$\\
&Quasar&5&$-$&$-$&$-$&$-$\\
6CE&Total&58&$-$&$-$&23&38\\
&LERG&8&$-$&$-$&4&7\\
&LERG?&11&$-$&$-$&6&8\\
&HERG&24&$-$&$-$&11&16\\
&HERG?&4&$-$&$-$&2&3\\
&Quasar&9&$-$&$-$&$-$&$-$\\
7CE&Total&74&$-$&6&36&54\\
&LERG&4&$-$&$-$&2&3\\
&NLRG&46&$-$&6&29&38\\
&BLRG&2&$-$&$-$&1&1\\
&Quasar&22&$-$&$-$&4&12\\
\hline
\end{tabular}
\caption[]{Table shows our detections for each sample in four $\textit{WISE}$ bands. LERG and HERG with question mark indicate that the classification is not certain. Many of the sources had upper-limit measurements in the longer $\textit{WISE}$ bands (12 and 22-$\mu$m). Among the samples, mostly NLRGs and LERGs have upper limits at long wavelengths.} {\label{samples}}
\end{table*}

\subsection{Consistency between $\textit{WISE}$ and $\textit{Spitzer}$}
\begin{figure}
\begin{center}
\scalebox{0.89}{
\begin{tabular}{c}
\centerline{\hspace{-1.0em}\includegraphics[width=10cm,height=10cm,angle=0,keepaspectratio]{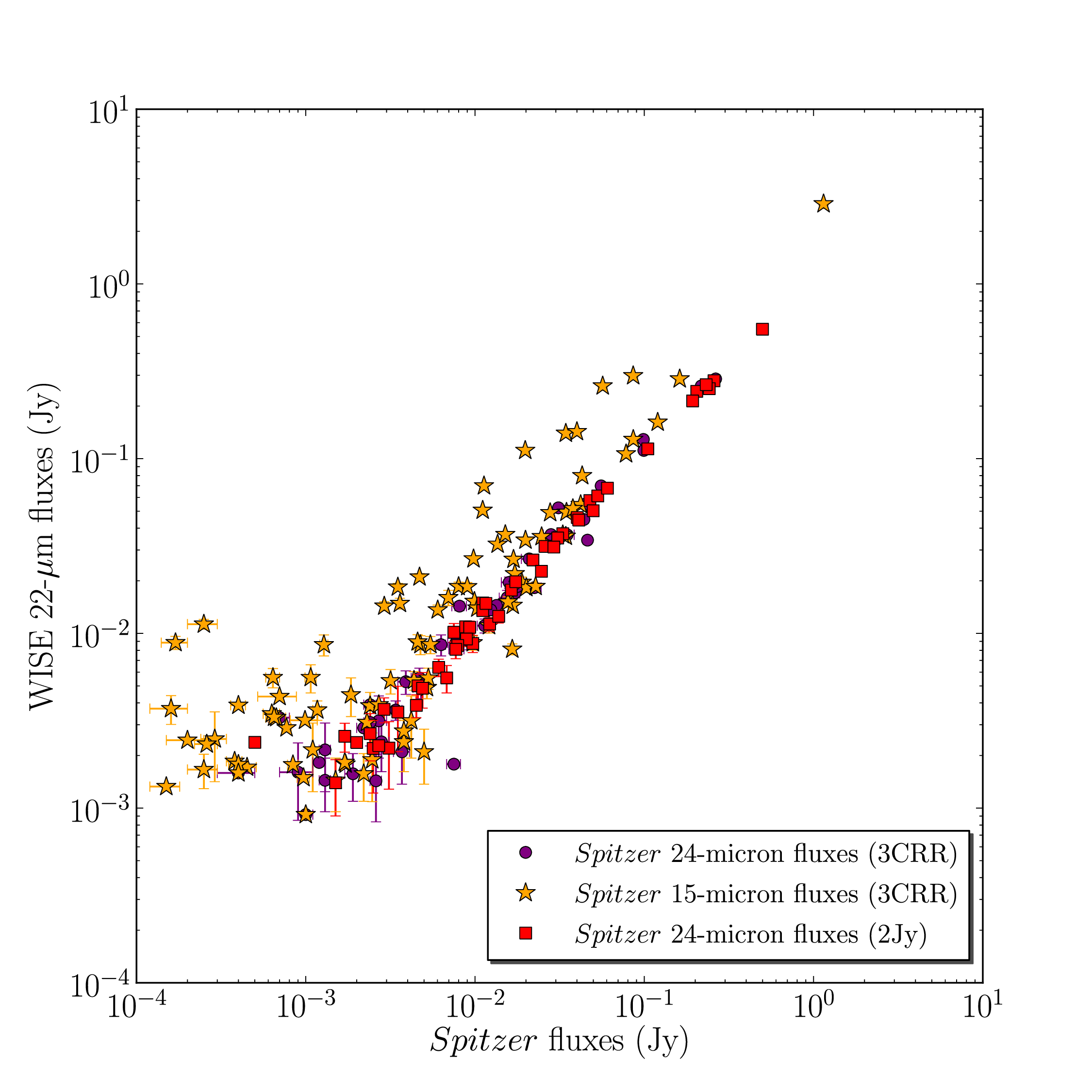}}\\
\end{tabular}}
\caption[.]{To show the consistency between $\textit{WISE}$ and $\textit{Spitzer}$ data we used sources in the 3CRR and 2Jy samples that have $\textit{Spitzer}$ measurements \citep{2006ref20,2007ref43,2008ref11,2009ref51}. Stars indicate 24-$\mu$m and circles 15-$\mu$m $\textit{Spitzer}$ fluxes of sources from the 3CRR sample, and squares indicate 24-$\mu$m $\textit{Spitzer}$ fluxes of the 2Jy sample sources. The 22-$\mu$m $\textit{WISE}$ magnitudes were plotted versus $\textit{Spitzer}$ magnitudes. A good agreement between the fluxes measured via $\textit{WISE}$ and $\textit{Spitzer}$ is clearly seen. \label{spitzer_wise}}
\end{center}
\end{figure}

Many sources in the samples considered here have mid-IR imaging or spectroscopy, taken by the $\textit{Spitzer}$ Space Telescope \citep{2004ref73}, therefore, we are able to check the consistency between $\textit{WISE}$ and $\textit{Spitzer}$ data. In order to do this, we used $\textit{Spitzer}$ measurements of 3CRR and 2Jy sources.  In the 3CRR sample, 92 sources have 15-$\mu$m rest-frame \citep{2006ref20,2007ref43,2009ref51} and 49 objects have 24-$\mu$m observed-frame, while all 2Jy objects (48) have 24-$\mu$m observed-frame $\textit{Spitzer}$ measurements. Our results can be seen in Figure \ref{spitzer_wise}. A comparison of the 24-$\mu$m
$\textit{Spitzer}$ and 22-$\mu$m $\textit{WISE}$ fluxes of both samples show excellent agreement between $\textit{Spitzer}$ and $\textit{WISE}$ data. There is also a good agreement between 15-$\mu$m (rest-frame) and 22-$\mu$m (lab-frame) fluxes although as expected, it shows a larger scatter in the correlation. 

\subsection{$\textit{WISE}$ luminosities}
 Before computing $\textit{WISE}$ luminosities in the four $\textit{WISE}$ bands, the 12-$\mu$m$-$22-$\mu$m spectral indices ($\textit{F$_{\nu}$}$\,$\alpha$\,$\nu^{-\alpha}$) were calculated. The mean index value (2.45, SD=0.88) corresponding to NLRGs, which should not be affected by contamination from direct quasar emission, of the 3CRR population where we have good detections in both bands was used for $\textit{K}$-corrections of the whole sample. The luminosities of the objects in the 12 and 22-$\mu$m (observed) bands were then calculated. The same process has been implemented for computing the 3.4 and 4.6-$\mu$m luminosities of the sources using the 3.4-$\mu$m$-$4.6-$\mu$m spectral indices (the mean index value is 0.14). $\textit{K}$-corrections were not derived using Spectral energy distributions (SED) because different components (old stellar population, torus and non-thermal emission\footnote{Although our objects are steep-spectrum sources, a minimal effect of non-thermal contamination may still be observed.}) are included in the mid-IR. However, it should be noted that this method of calculating the K-correction in 3.4-$\mu$m band may not provide the best results because it does not take into account any curvature below 3.4-$\mu$m which is expected for radio galaxies; thus 3.4$\mu$m luminosities should be treated with caution. 151-MHz luminosities were extrapolated for the 3CRR and 2Jy samples using the radio spectral indices and 178-MHz or 408-MHz flux density measurements from the catalogues. 5-GHz core flux densities regarding the 3CRR and 2Jy objects in the catalogues are used. 

%%%%%%%%%%%%%%%%%%%%%%%%%%%%%%%%%%%%%%%%%%%%%%%%%%%%%%%%%%%%%%%%%%%%%%%

\section{Results}

\subsection{Colour$-$colour diagrams}
\begin{figure*}
\begin{center}
        %\resizebox{0.497\hsize}{!}{\includegraphics{wise1_z.eps}}
        %\resizebox{0.497\hsize}{!}{\includegraphics{wise2_z.eps}} 
        \resizebox{0.497\hsize}{!}{\includegraphics{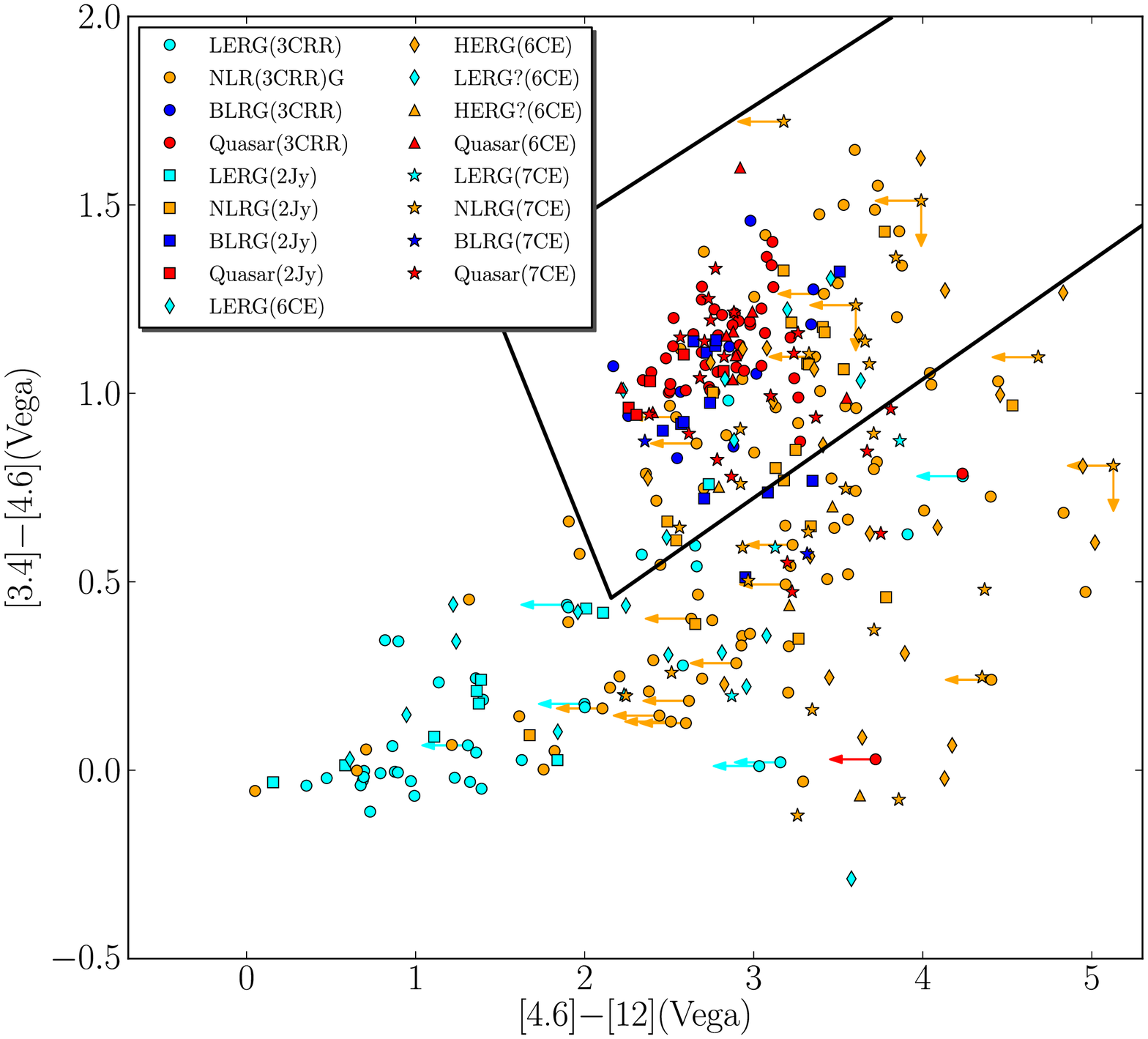}}
        \resizebox{0.497\hsize}{!}{\includegraphics{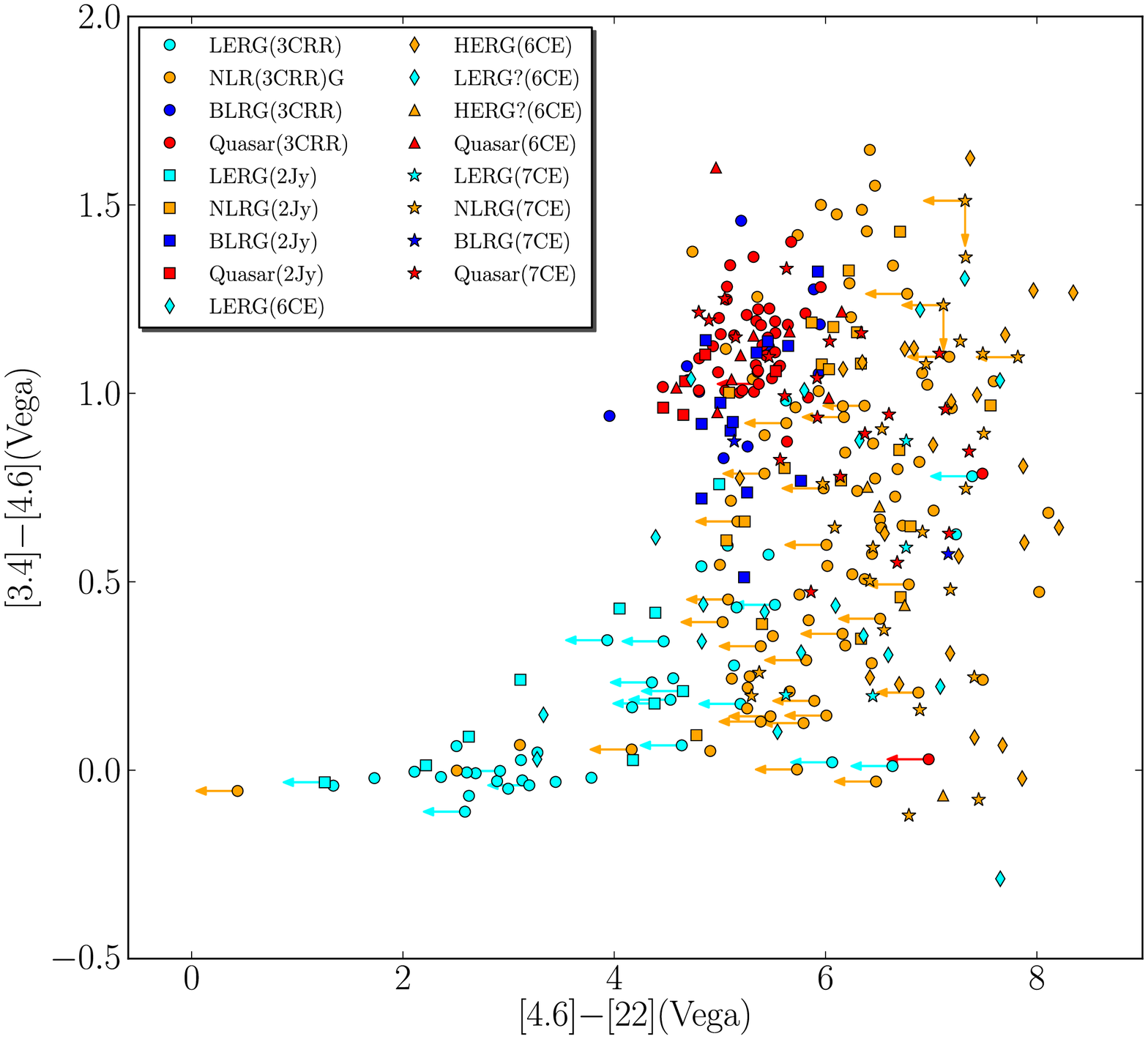}}
        \caption{$\textit{WISE}$ colour-colour diagrams of all objects in the combined sample. In the left plot for comparison we also show the three-band AGN wedge ($W2-W3\geq 2.517$, $W1-W2>0.315\times (W2-W3)-0.222$ and  $W1-W2<0.315\times (W2-W3)+0.796$) with black solid box defined by \citet{2012ref100}. Upper limits are indicated as arrows.}
\label{colour_colour}
\end{center}\vspace{-1em}
\end{figure*}

Various methods have been developed for separating AGN from normal galaxies in the mid-IR \citep[e.g.][]{2004ref98,2005ref99}.  In particular, $\textit{WISE}$ colours have been utilised to select AGN \citep[e.g.][]{2010ref104,2011ref101,2012ref100,2012ref102,2012ref115,2013ref103}. Before carrying out our main quantitative analysis, we investigated the positions of our objects in $\textit{WISE}$ colour$-$colour diagrams. Figure \ref{colour_colour} shows colour-colour plots produced by using $\textit{WISE}$ $W1$, $W2$, $W3$ (3.4, 4.6, 12-$\mu$m$-$Vega magnitudes)  and $\textit{WISE}$ $W1$, $W2$, $W4$ (3.4, 4.6, 22-$\mu$m$-$Vega magnitudes). Upper limits are indicated as arrows in the plots. 

The $W1$,$W2$ and $W3$ colours are widely used for selecting AGN \citep[e.g.][]{2012ref100}. For comparison in Figure \ref{colour_colour} (left) we show the AGN wedge with the black solid box defined by \cite{2012ref100}.
 It can be seen in Figure \ref{colour_colour} that these widely used colour cuts are not reliable for selecting all types of AGN. All LERGs and almost half of NLRGs are omitted if these colour cuts are used. This suggests that $\textit{WISE}$ colour cuts should be used with care for selecting AGN: in particular the fact that many NLRGs are not selected suggests that such colour cuts may be biased against heavily obscured type-2 AGN.

In both plots quasars and BLRGs have similar colours. NLRGs occupy a range of colours, which is suggestive of varying amounts of quasar contamination in the $\textit{WISE}$ near-IR bands. Using $W4$ instead of $W3$ leads to a better separation of LERGs and NLRGs. To determine the separation quantitatively, the best separation line was chosen at $W2-W3=2$ and $W2-W4=4.9$. The number of LERGs and NLRGs were counted in each side. For $W2-W3<2$ the number of LERGs is 40 and of NLRGs is 16, and for $W2-W3>2$ there are 30 LERGs and 159 NLRGs. For $W2-W4<4.9$ there are 41 LERGs and 10 NLRGs, the number of LERGs is 29 and of NLRGs is 165 for $W2-W4>4.9$. Thus, the $W2-W4$ selection is slightly better at rejecting NLRGs from the LERG region. This shows that the effect of the torus is stronger in 22-$\mu$m band, as expected.
 
%%%%%%%%%%%%%%%%%%%%%%%%%%%%%%%%%%%%%%%%%%%%%%%%%%%%%%%%%%%%%%

\subsection{Investigation of the old stellar population}
\begin{figure*}
\begin{center}
\scalebox{0.89}{
\begin{tabular}{c}
\centerline{\hspace{-0.9em}\includegraphics[width=10cm,height=10cm,angle=0,keepaspectratio]{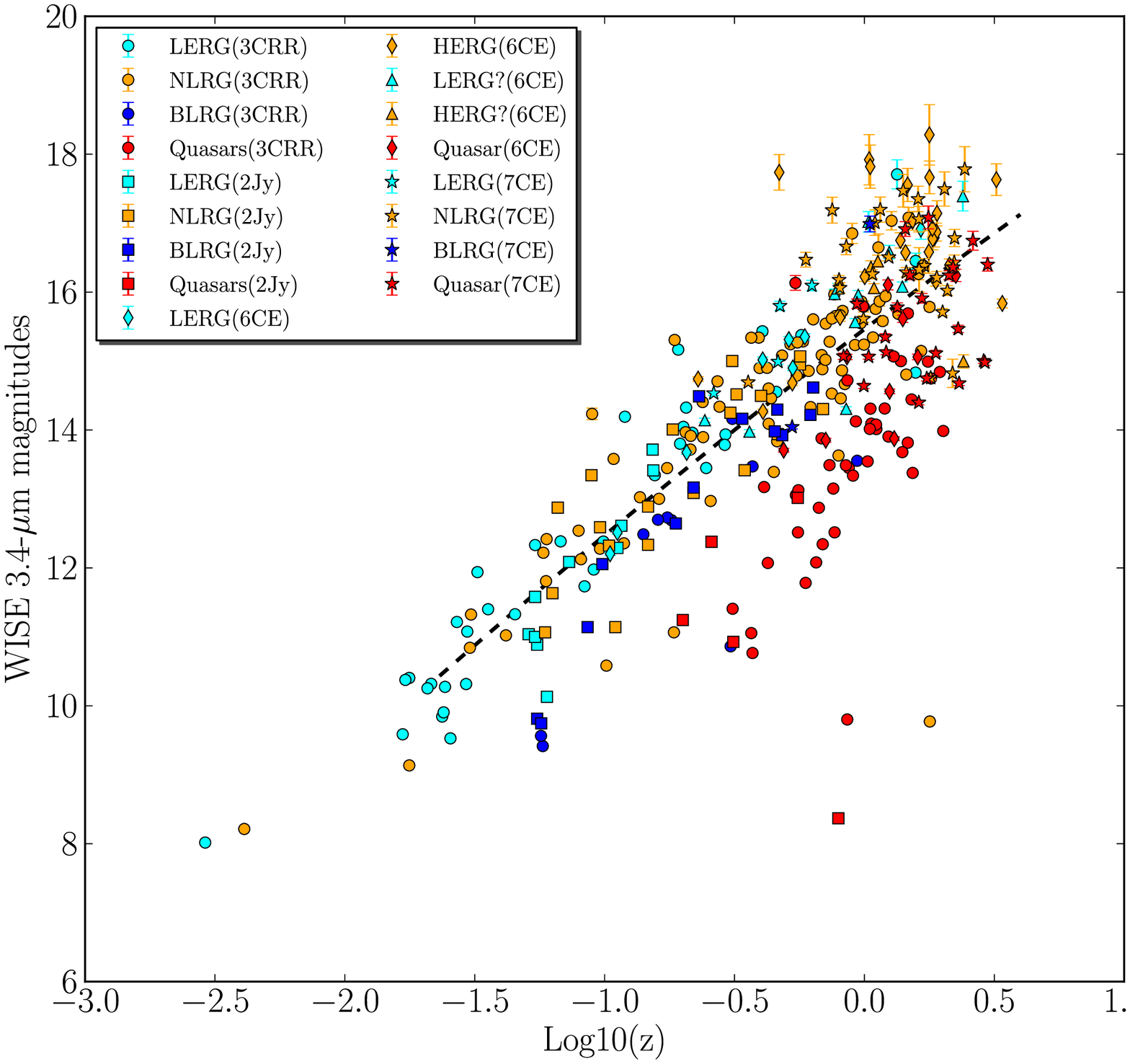}}\\
\end{tabular}}
\caption[Plot of 3.4$\mu$m magnitudes versus redshift. A black dashed line shows the best fit.]{Plot of 3.4$\mu$m magnitudes versus redshift. A black dash line shows the best fit. \label{K_z}}
\end{center}
\end{figure*}

Since the near-IR emission from radio galaxies is dominated by the old stellar population of the host galaxies, studies of radio objects over a wide redshift range in the near-IR provide insights into the evolution of stellar components underlying such objects. The distribution of 3.4-$\mu$m magnitudes versus redshift is shown in Figure \ref{K_z}. In this figure a relationship between near-IR band and redshift; an increase in magnitude with redshift, which is identical to the $K$-z relation of radio galaxies seen in previous studies of various radio samples \citep[e.g][]{1982ref58,1997ref6,2001ref57,2003ref15,2010ref60}. The relation between 3.4-$\mu$m and redshift is modelled by fitting a second-order polynomial (15.46 + 2.85\,log$_{10}z$-0.13(log$_{10}z$)$^{2}$) which provided the best fit to the data. Only NLRGs and LERGs were used in the fits because quasars/BLRGs are contaminated by non-stellar quasar emission. A tight correlation between K magnitudes  and redshift has been interpreted as showing that the radio galaxies are a homogeneous population associated with giant elliptical galaxies containing old stellar populations at lower redshifts. They formed the majority of their stars at high redshifts ($z\ga3$, but see \cite{1998ref59,1999ref83}) and evolved passively thereafter \citep[e.g.][]{1989ref46,2000ref85,2001ref57}. Our results provide no motivation for doubting this conclusion. 

\begin{figure*}
\begin{center}
        \resizebox{0.497\hsize}{!}{\includegraphics{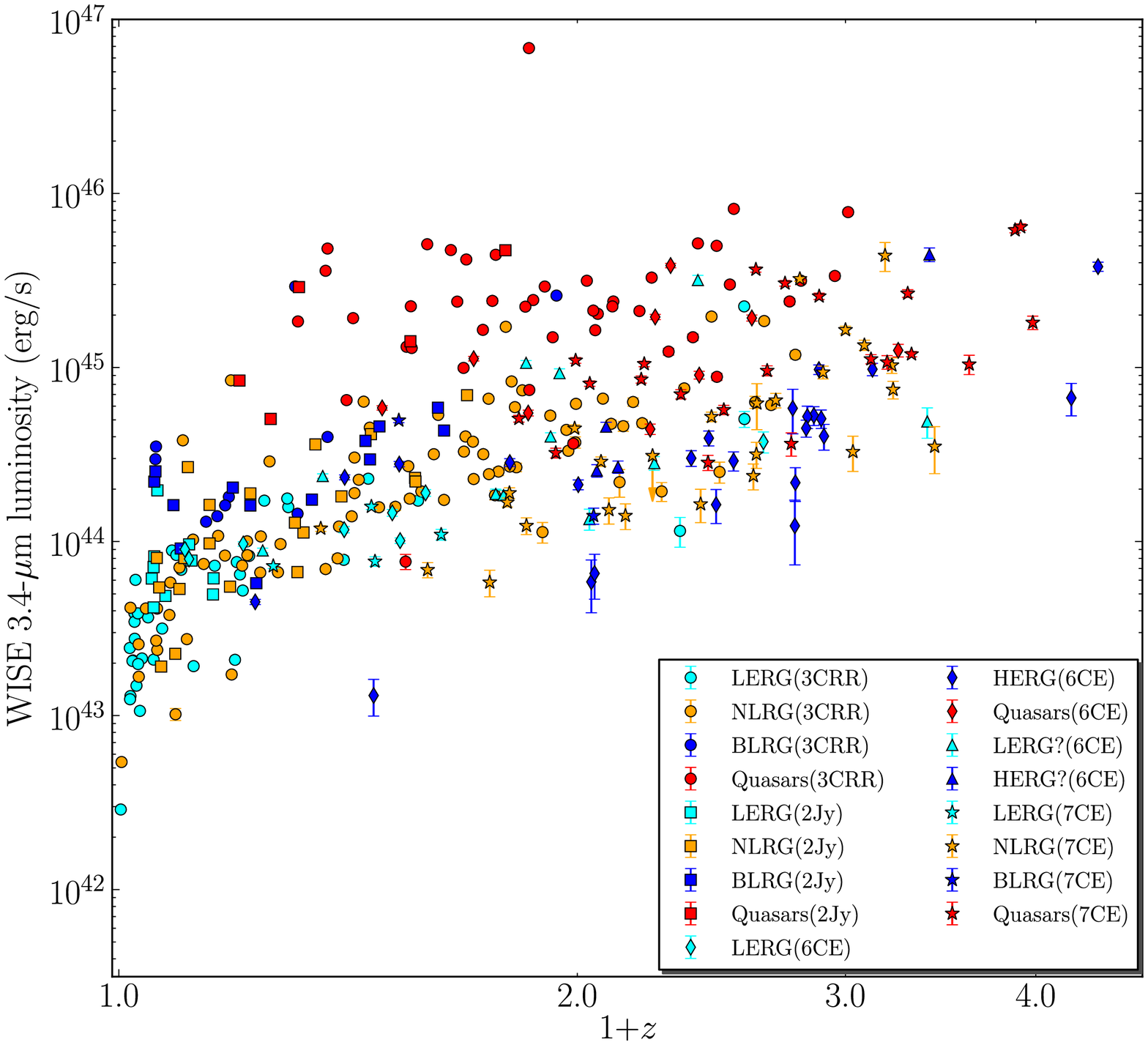}}
        \resizebox{0.497\hsize}{!}{\includegraphics{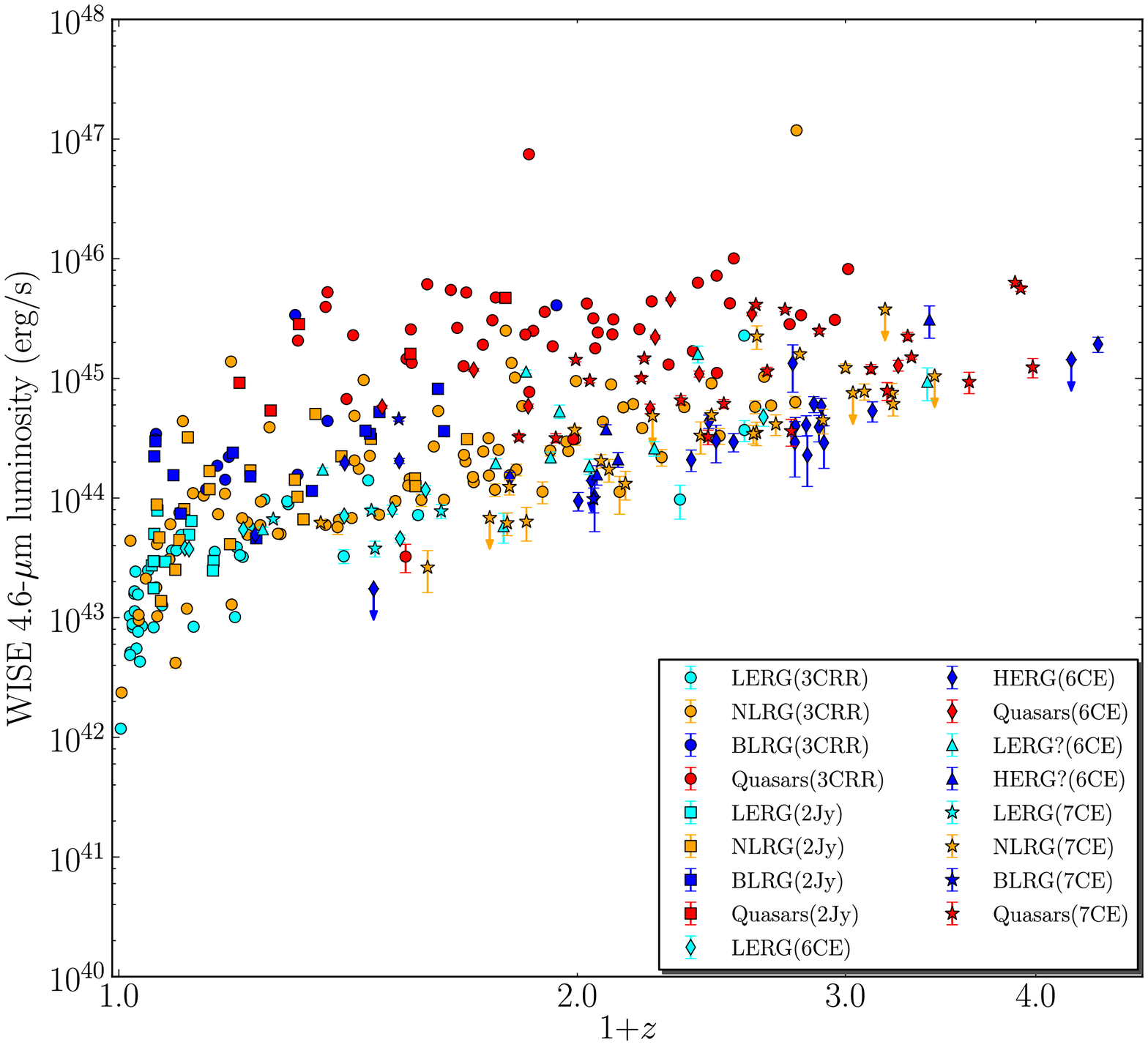}} 
        %\resizebox{0.497\hsize}{!}{\includegraphics{wise1_z_k2.eps}}
        %\resizebox{0.497\hsize}{!}{\includegraphics{wise2_z_k2.eps}}
        \caption{Plot of 3.4$\mu$m luminosity (left) and 4.6$\mu$m luminosity (right) versus redshift. Upper limits are indicated as arrows. }
\label{wise1_2}
\end{center}\vspace{-1em}
\end{figure*}

In Figure \ref{wise1_2} we also show the 3.4-$\mu$m luminosity and 4.6-$\mu$m luminosity versus redshift of the sources. 3CRR and 2Jy sources display almost identical near-IR luminosities. On the other hand, 6CE and 7CE objects lie close to each other and they have fainter host galaxies than the 3CRR sample at high redshift ($z\gtrsim 0.7$). To quantify this, we calculated the ratio of median fluxes of the samples for different redshift bins. The ratio of median fluxes for the 3CRR and 2Jy samples' galaxies in the 3.4-$\mu$m waveband is 0.7 (0.08) for $z=0-0.7$. This ratio for the 6CE and 7CE samples is 1.8 (0.8) for $z=1-1.517$. If the 3CRR and 6CE samples are considered the flux ratio is 1.7 (0.3) for the same redshift bin. A similar trend was seen in other studies \citep[e.g.][]{1997ref6,1998ref59,2001ref57,2002ref84,2003ref15,2004ref107}.

 A possible explanation for this is that 3CRR sources are hosted by more massive systems (and proportionally more luminous) compared to the 6CE and 7CE radio galaxies \cite[e.g.][]{1998ref59}. The jet power of a radio source is expected to be determined by the mass of the black hole in the centre of the source, the matter that accretes on to it and the efficiency of the accretion. Thus, powerful radio sources are more likely to host high mass black black holes. As many observational studies have shown \citep[e.g.][]{1995ref64,1998ref65,2001ref108} that the central black hole masses are roughly in proportion with the mass of host galaxies, 3CRR and 2Jy galaxies would be expected to tend to reside in more massive galaxies which are brighter at 3.4$\mu$m than the 6CE and 7CE galaxies.

Regarding different classes, we see that in each sample BLRGs have similar distributions to NLRGs in both near-IR bands. They could be either intrinsically weak AGN or reddened quasars. LERGs and NLRGs also show similar near-IR luminosities over the redshift range. The 6CE sample is classified mainly as LERGs and HERGs so HERGs will include some NLRGs and BLRGs. In both plots, we see that HERGs (6CE) mostly appear close to NLRGs (6CE) which is not surprising since they will most likely be narrow-line objects. Quasars are more luminous at both near-IR wavelengths in comparison with radio galaxies in each sample. This is most plausibly due to non-stellar quasar emission that contributes to the near-IR band.

\subsection{ The possibility of a starburst contribution?}
In Figure \ref{unification} we show the mid-IR luminosity distribution (observed frame 12-$\mu$m luminosity and 22-$\mu$m luminosity) of different class of sources as a function of redshift. It is important to consider whether any of this mid-IR emission could originate from warm dust heated by star formation. Therefore, we investigated any contribution of a starburst at the 22-$\mu$m waveband. In the most extreme case the integrated IR luminosity of a starburst is around 10$^{46}$ erg s$^{-1}$ \citep{2011ref117}. We used the  relation log($L_{\rm IR}$) = 1.02+0.972$\times$log($L_{\rm 12\mu m}$) given by \cite{2005ref118} to estimate the luminosity at 22-$\mu$m band. This value is around 10$^{45}$ erg s$^{-1}$. Any object brighter than this cannot be a starburst. This allows us to rule out this possibility for most of our objects. Most of the remaining sources are LERGs and NLRGs, but these do not present starburst colours on the colour-colour diagrams of Figure \ref{colour_colour}. Starbursts would occupy the right bottom corner. LERGs have redder colours and sit in the bottom left region on the plots. Furthermore, a continuum between the properties of sources above and below 10$^{45}$ erg s$^{-1}$ is apparent in the 12- and 22-$\mu$m luminosity versus redshift plots. We conclude that star-formation activity cannot contribute significantly to the observed mid-IR emission and continue on the assumption that it is related primarily to AGN activity.

%%%%%%%%%%%%%%%%%%%%%%%%%%%%%%%%%%%%%%%%%%%%%%%%%%%%%%%%%%%%%%%
\subsection{Quasar-radio galaxy unification}
\begin{figure*}
\begin{center}
\scalebox{0.89}{
\begin{tabular}{c}
\includegraphics[width=12cm,height=12cm,angle=0,keepaspectratio]{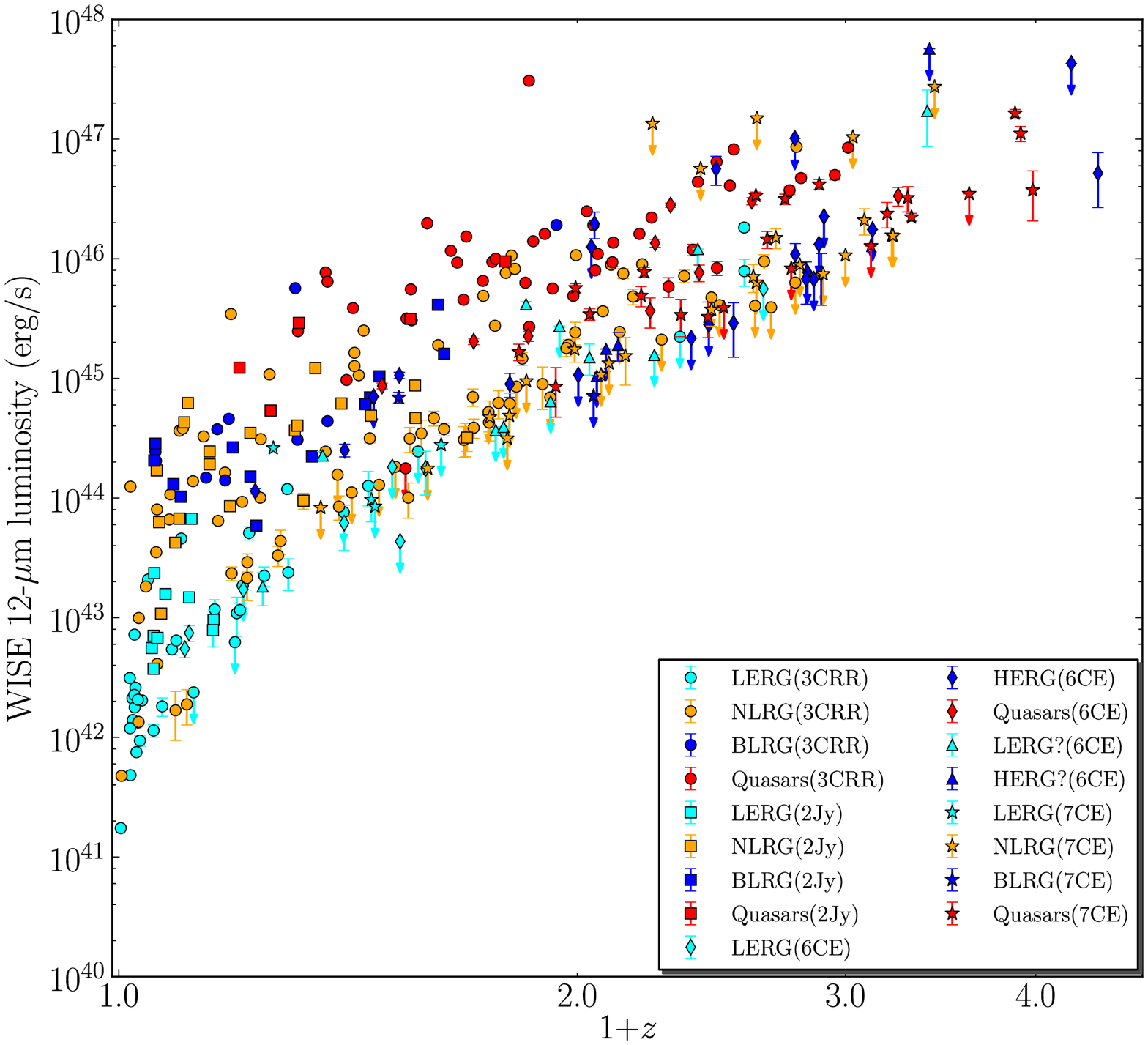}\\
\includegraphics[width=12cm,height=12cm,angle=0,keepaspectratio]{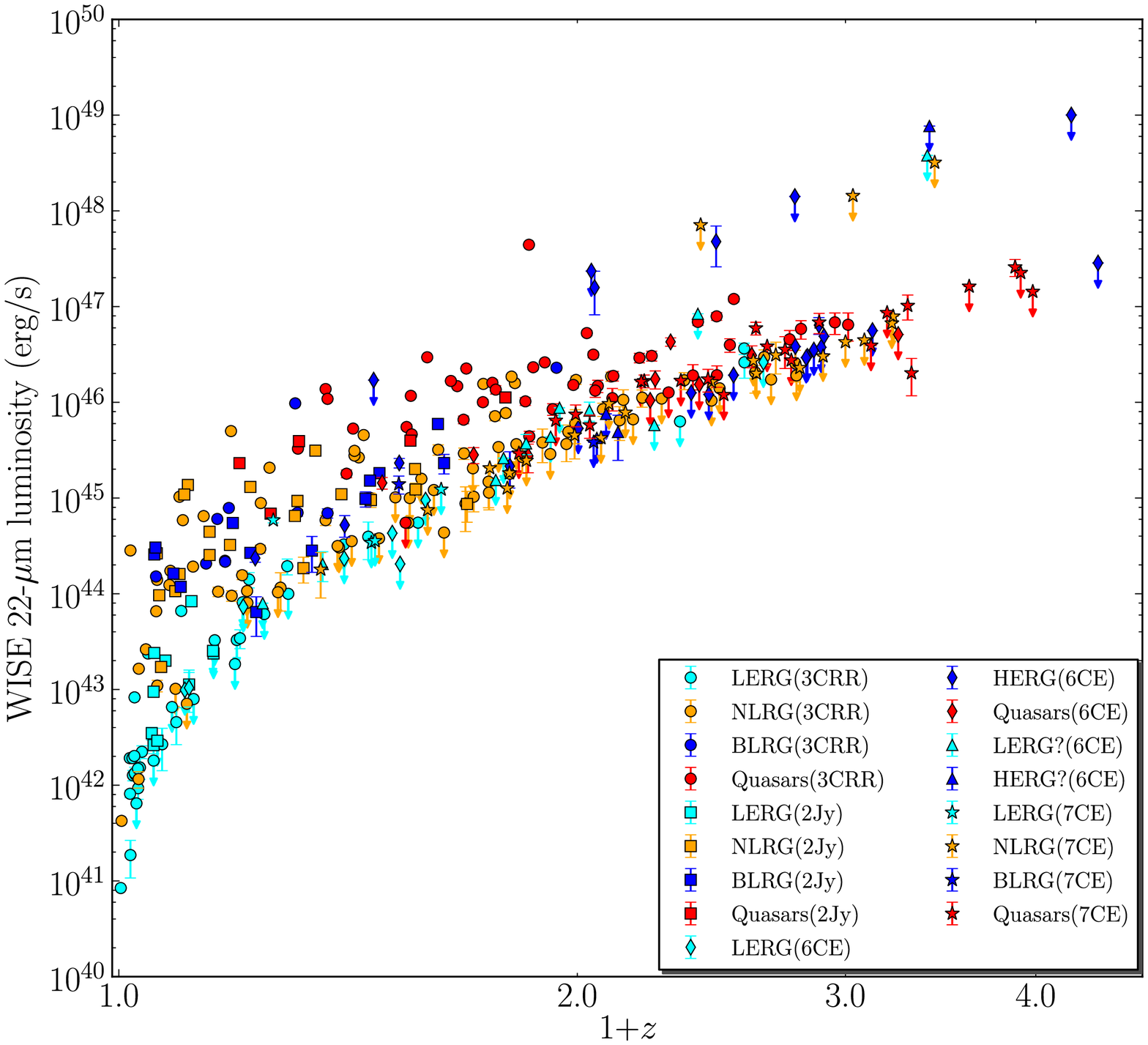}\\
\end{tabular}}
\caption[]{Plot of 12$\mu$m luminosity (top) and 22$\mu$m luminosity (bottom) versus redshift. Upper limits are indicated as arrows. \label{unification}}
\end{center}
\end{figure*}

In standard AGN models, optical/UV emission obscured by dusty structures around the accretion disc is re-radiated in the mid-IR. Thus, hidden quasars can be inferred by means of mid-IR observations. It can be clearly seen in Figure \ref{unification} that there is more scatter in the 12$\mu$m luminosities than in the 22$\mu$m luminosities. The old stellar population has a slight contribution at 12 $\mu$m, but possibly more importantly PAHs have a strong effect at this wavelength \citep[e.g.][and references therein]{2006ref116}, thus the higher scatter may be attributed to this. 

A comparison of the mid-IR properties of different classes of AGN reveals that quasars have higher mid-IR luminosities with respect to NLRGs. To quantify any excess in quasar mid-IR emission, median values of 12-$\mu$m and 22-$\mu$m luminosities in 4 redshift bins were obtained for both quasars and NLRGs. Since we have upper limits, to calculate median values we use survival-analysis statistics which are used for data sets with censored data. The median values were derived using the ASURV \citep{1985ref111,1985ref109,1986ref110} computer package which is uses the Kaplan-Meier estimator. Errors for the estimates of the median values were obtained using the bootstrapping technique. The ratios of the median values then quantify this excess emission in quasars. Calculated median values with their errors, selected redshift bins and the ratios are given in Table \ref{results_unification}. LERGs have lower mid-IR luminosities than the other class of objects, as seen in both plots. Although BLRGs tend to have higher mid-IR luminosities than NLRGs, an overlap between NLRGs and BLRGs is seen. To show this quantitatively median values of the 22-$\mu$m fluxes in 3 redshift bins for both BLRGs and NLRGs were obtained. These values were used to calculate the flux ratios. The ratios are 2.5 (0.4), 0.8 (0.07) and 1.0 (0.05) for the redshift bins $z=0.05-0.75$, $z=0.75-1.45$ and $z=1.45-2.15$, respectively.

In all redshift bins, quasars exhibit stronger mid-IR emission than NLRGs. Since re-radiated UV-optical emission from AGN dominates the 22-$\mu$m flux, ratios of the median values regarding the objects for this waveband can give the best indication of the difference in the mid-IR emission between quasars and radio galaxies. In the first redshift bins quasars are about 10 times more luminous than radio galaxies. The ratio decreases towards higher redshifts although quasars still have stronger behaviour in their mid-IR luminosities ($\approx$ 3-2 times more luminous than radio galaxies). The decrease in the ratio towards high redshifts does not appear to be a simple effect of redshifting non-evolving SEDs of quasars and radio galaxies. Obtaining the distribution of SEDs as a function of luminosity and redshift would allow us to study this particular trend in more detail, but this is beyond the scope of this paper. 

A significant difference ($\gtrsim 2$) in the mid-IR magnitudes of quasars and radio galaxies was also reported by several studies \citep[e.g.][]{1992ref69,1994ref56,2005ref71,2007ref43,2011ref121}. Other authors \citep[e.g.][]{2001ref21,2004ref70} used small samples and found a slight difference between the properties of quasars and galaxies in the IR bands. Various interpretations (such as torus anisotropy, effects of the environments) were used to explain this. \cite{2009ref10} obtained no difference between the mid-IR luminosities of the broad- and narrow-line objects from the 2Jy sample. However, although they used a complete sample, it is small in size. Our work is the first to use large, complete samples with coherent mid-IR data and good coverage of the redshift range where quasars and radio galaxies coexist in large numbers.

%Our findings with regards to the unification of quasars and radio galaxies are in agreement with various previous studies' results \citep[e.g.][]{1992ref69,1994ref56,2005ref71,2007ref43}.  Although they used complete samples, these are small in size.} 
%The contradictory results of similar studies obtained in the literature may be due to the incomplete samples and incoherent mid-IR data used.

Different torus models have been proposed to explain different properties of quasars and radio galaxies \citep[e.g.][]{1992ref75,1993ref76,2002ref77,2008ref78,2008ref79,2005ref97,1991ref91,1998ref92,2003ref93} and anisotropic emission due to the torus is expected in all suggested models. In order to evaluate our results in terms of different torus models we first calculated the critical angles ($\theta_{crit}$) which is the angle to the line of sight separating quasars and radio galaxies. To do that, as described in \cite{1989ref17} we used the probability function of finding a source within an angle to the line of sight for a randomly distributed set of sources and the number of quasars and radio galaxies found in a given redshift bin. The characteristic angles (the expected angle to the line of sight) for quasars ($\theta_{\rm Q}$) and radio galaxies ($\theta_{\rm RG}$) were computed for each redshift bin. These results are also shown in Table \ref{results_unification}. 

We then examined our results with regard to the predictions of well-known torus models such as torus with smooth density distribution \citep{2005ref97} and torus with clumps \citep{2008ref79}. Our results for both 12 and 22-$\mu$m wavebands are in agreement with the predictions of smooth torus models which show a relatively strong effect of anisotropy compared to clumpy torus models. For instance, there is a higher difference in the luminosities of quasars and radio galaxies at lower redshifts compared to the higher redshifts. Figure 8 in \cite{2005ref97} shows an inclination angle study for a range of wavelengths considering a smooth torus model; according to this figure, for the first two redshift bins, we expect to see a strong anisotropy in the mid-IR luminosity for calculated characteristic angles of quasars and radio galaxies at a corresponding wavelength (from the angles around 30$^{\circ}$ to 70$^{\circ}$ approximately a factor of four decrease is expected at 11 $\mu$m (rest-wavelength)). A similar study for clumpy torus models \citep[][their Figure 10]{2008ref79} shows only a slight difference in the mid-IR magnitudes for expected angles of quasars and radio galaxies (the difference in the mid-IR magnitudes between the angles around 30$^{\circ}$ and 70$^{\circ}$ is a factor of 1.2 at 12$\mu$m (rest-wavelength)). As previously mentioned there are three components that contribute to the mid-IR: re-radiation from torus; emission from the old stellar population and non-thermal contamination that can be seen in quasars/BLRGs. Any effect from an old stellar population is expected to be the same for both quasars and NLRGs. In clumpy torus models, substantial non-thermal contamination in quasars/BLRGs would have to be present to explain our results.

%a low object with starburst characteristics would have a total IR luminosity around 10$^{9}$ L$_{\odot}$. We have used the far-IR$-$radio relation $q$ = log$_{10}(S_{\rm FIR}/3.75\times10^{12}\times S_{\rm 1.4}$) given by \cite{1985ref119} and a standard value of q of a starburst to estimate the luminosity at 1.4-GHz. This value was then used to obtain 151-MHz radio luminosity which is around 10$^{42}$ erg s$^{-1}$. Almost all LERGS have radio luminosities lower than this value so they cannot have starbursts.} 

\begin{table*}
\begin{tabular}{cccccccccc}
\hline
Redshift bins&Class&N&3.4$\mu$m&4.6$\mu$m&12$\mu$m&22$\mu$m&$\theta_{crit}$&$\theta_{Q}$&$\theta_{RG}$\\
&\& Ratio&&erg/s&erg/s&erg/s&erg/s&&&\\
\hline
0.2$<$z$<$0.621&Quasars&15 &1.365 (0.350)  &1.531 (0.432)  &2.983 (0.6285)  &3.951 (0.913)   &&32.2\\
                 &NLRGs&49 &0.142 (0.017)  &0.090 (0.015)  &0.137 (0.0885)  &0.239 (0.167)   &&&70.4\\
                   &Ratio& &9.612 (2.720)  &17.011 (5.623) &21.773 (14.794) &16.531 (12.199) &49\\
\hline

0.621$<$z$<$1.00&Quasars&20&1.645 (0.633)  &1.913 (0.5325) &6.298 (1.329)   &10.232 (2.538)  &&36\\
                  &NLRGs&38&0.270 (0.042)  &0.154 (0.0305) &0.437 (0.112)   &1.433 (0.171)   &&&73\\
                    &Ratio&&6.092 (2.533)  &12.422 (4.243) &14.411 (4.797)  &7.140 (1.965)   &55\\
\hline 

1.00$<$z$<$1.517&Quasars&23&1.566 (0.458)  &1.738 (0.424)  &8.852 (1.671)   &17.077 (2.233) &&45.4&\\
                  &NLRGs&25&0.264 (0.036)  &0.288 (0.061)  &1.553 (0.510)   &4.951 (1.288)   &&&80.1\\
                    &Ratio&&5.931 (1.881)  &6.034 (1.957)   &5.699 (2.160)  &3.449 (1.004)  &70\\
\hline 

1.517$<$z$<$1.96&Quasars&11&2.782(0.397)   &3.231 (0.374)  &35.475 (3.866)  &42.618 (8.610)  &&44.8\\
                  &NLRGs&21&0.556(0.052)   &0.431 (0.094)  &2.826 (1.874)   &16.106 (3.547)  &&&79.6\\
                    &Ratio&&5.003 (0.856)  &7.496 (1.850)  &12.553 (8.043)  &2.646 (0.790)   &69\\
\hline 
\end{tabular}
\caption[]{Median values of the 3.4-$\mu$m, 4.6-$\mu$m, 22-$\mu$m and 12$\mu$m luminosities were derived using the ASURV \citep{1985ref111,1985ref109,1986ref110} computer package for four different redshift bins. Redshift bins are shown in column 1, N indicates number of sources included to calculate median values, shown in column 3. Classes and ratios computed are given in column 4, 5, 6, 7 for the 3.4-, 4.6-, 12- and 22-$\mu$m luminosities respectively. Errors in calculations are given in parentheses. Column 8 shows the critical angles calculated and characteristic angles found for quasars as well as radio galaxies are given column 9 and 10.} {\label{results_unification}}
\end{table*}

%%%%%%%%%%%%%%%%%%%%%%%%%%%%%%%%%%%%%%%%%%%%%%%%%%%%%%%%%%%%%%%%%%

\subsection{Dichotomy for LERGs and HERGs}
Our primary aim in this section is to investigate how LERGs can fit into unification models and to what level mid-IR radiation can aid us in classifying LERGs and HERGs. The low-frequency radio luminosity is related to the time-averaged jet kinetic power but also the age of the source and to the properties of the external environment \citep[e.g.][]{2013ref82}. The mid-IR luminosity gives information about the torus emission (if it is present) and can be considered as a proxy for the intrinsic AGN luminosity \citep[e.g.][]{2011ref114}. There may be some contamination, such as emission from the old stellar population and any possible anisotropy in the torus emission, as mentioned in the previous section. There can also be contributions from jet related non-thermal emission, Doppler boosted due to low inclinations (mainly in quasars and BLRGs).

\begin{figure*}
\begin{center}
        \resizebox{0.497\hsize}{!}{\includegraphics{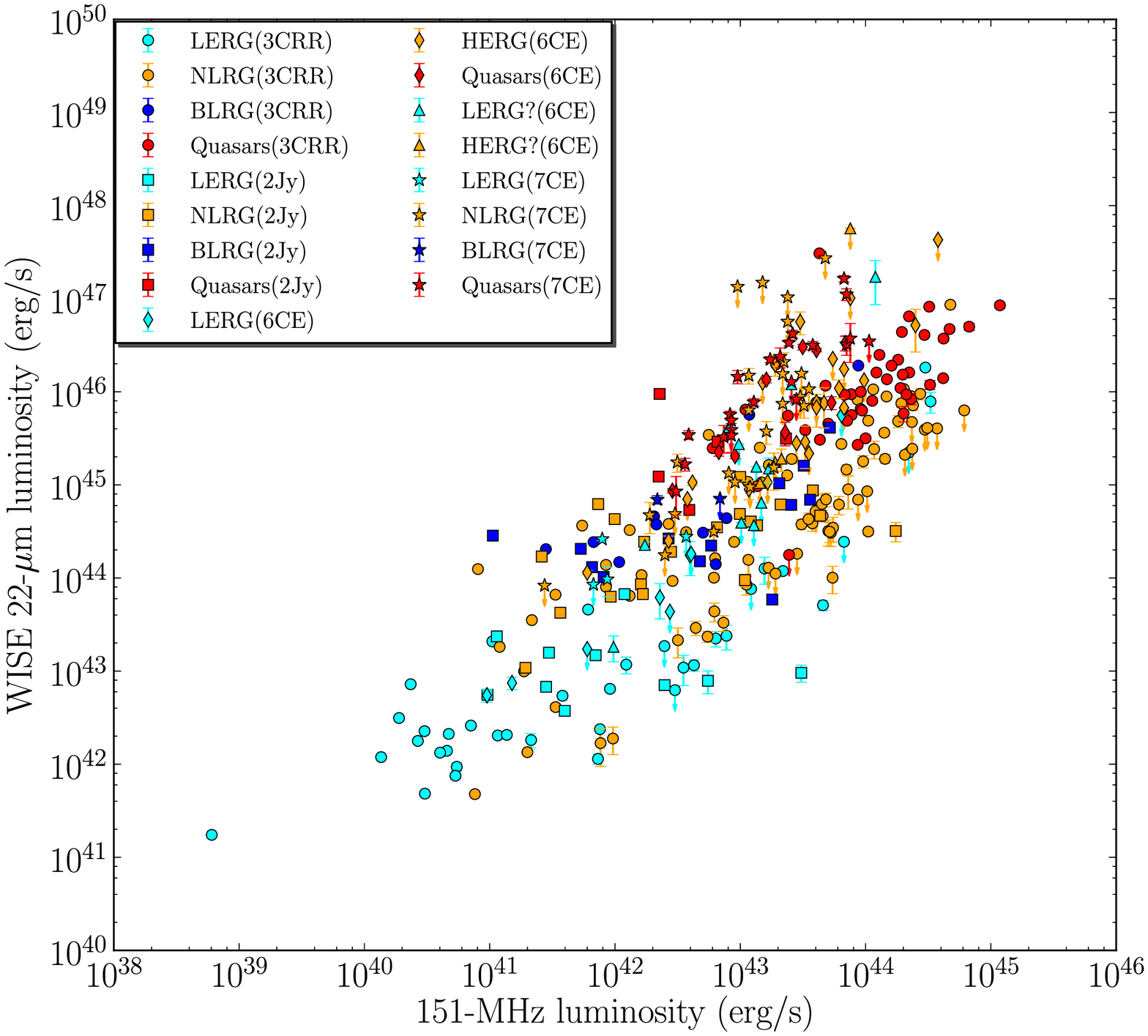}}
        \resizebox{0.497\hsize}{!}{\includegraphics{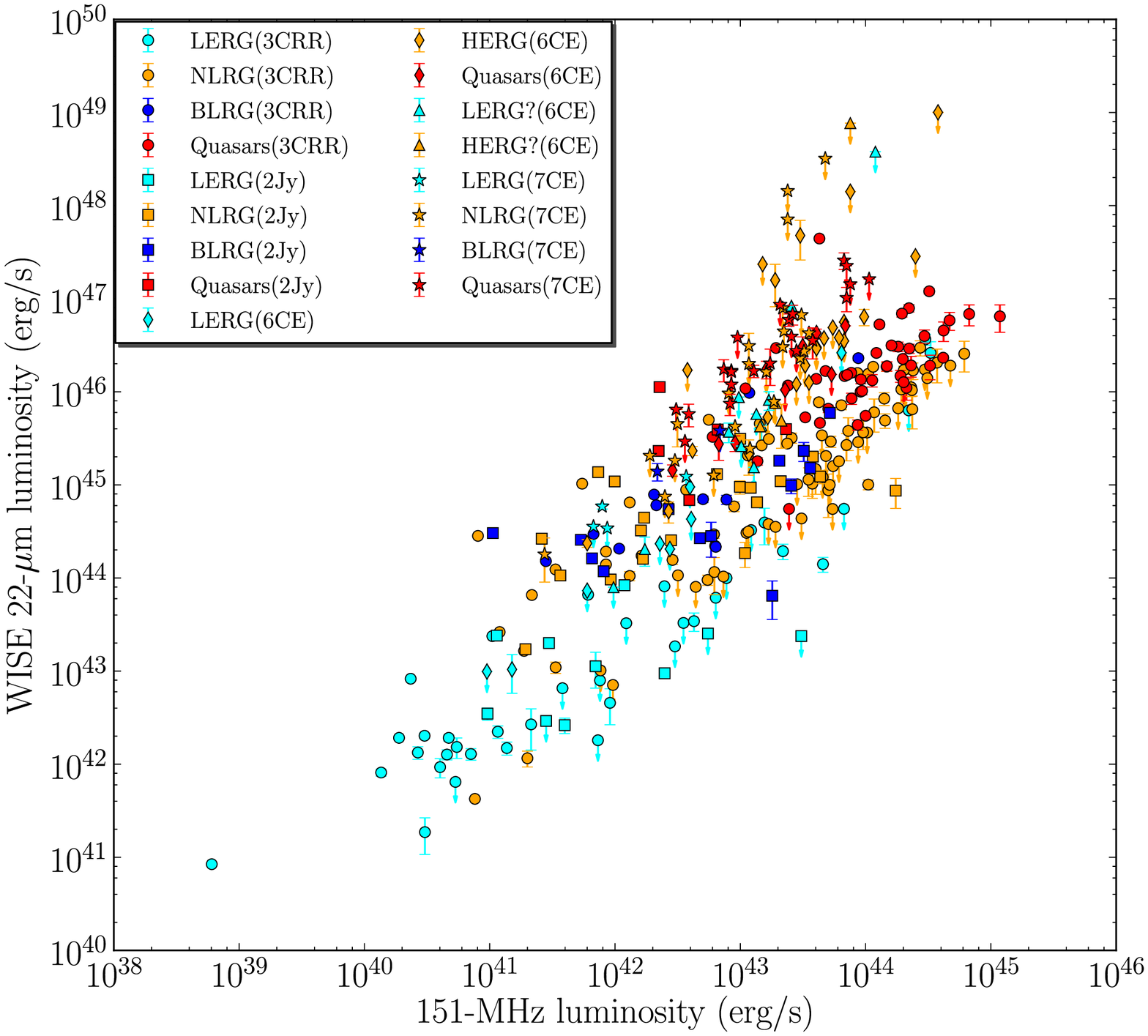}}
        \caption{Plot of 12$\mu$m luminosity (right) and 22$\mu$m luminosity (left) versus 151-MHz radio luminosity. Upper limits are indicated as arrows.}
\label{rad_ir}
\end{center}\vspace{-1em}
\end{figure*}

In Figure \ref{rad_ir} we present the mid-IR luminosity versus low-frequency radio luminosity (at 151-MHz) for the radio sources in our samples. In both plots we see a correlation between the 151-MHz and mid-IR luminosities. This is expected as presumably both luminosities are roughly isotropic indicators of AGN nuclear activity. The 12$\mu$m plot has higher scatter than the 22$\mu$m one; this is ascribable to strong PAH features and emission from an old stellar population which still contributes at 12-$\mu$m band. Further examination of each of the samples in both plots shows that there is a scatter in the correlation due to selection biases. The 3CRR/2Jy objects have high radio luminosities in comparison to 6CE/7CE objects for a given mid-IR luminosity. The 6CE/7CE objects exhibit lower jet powers in comparison to the 3CRR/2Jy sources for a chosen radiative power (For a given 22-micron luminosity  (L$_{22 \mu m}$ = 10$^{45}$ erg s$^{-1}$) the corresponding 151-MHz luminosity for the 3CRR sample is 3$\times$10$^{43}$ erg s$^{-1}$, this value is 4$\times$10$^{42}$ erg s$^{-1}$ for the 6CE sample and 1$\times$10$^{42}$ erg s$^{-1}$ for the 7CE sample.). This suggests that we may not necessarily see a one-to-one correlation between radiative power and jet power \citep[e.g. Mingo et al. 2013;][]{2006ref20,2011ref95,2011ref114}. 3CRR and 2Jy objects, which are the most radio-luminous sources in the Universe in their redshift range, are selected to have highest radio luminosity for a given AGN power.
% LERGs occupy a certain region of parameter space in the radio$-$mid-IR plane (we will only consider 22-$\mu$m luminosity $-$ 151-MHz luminosity since this is the best AGN power indicator) in comparison to HERGs.

Considering the LERG/HERG division, the correlation between the low-frequency radio luminosity and mid-IR luminosity is very clear for HERGs. However, this correlation disappears for LERGs. We have used a quantitative test of partial correlation that takes into account censored data \citep{1996ref80}. The results of this correlation analysis can be found in Table \ref{partial_corr}. We performed the correlation analysis between 22-$\mu$m and 151-MHz luminosity for all samples as well as for LERGs and HERGs separately. These results confirm the physical relationship between mid-IR and radio luminosities seen for HERGs (the strength of the correlation is given by $\tau$/$\sigma$=6.08). On the other hand, the relationship is not significant for LERGs ($\tau$/$\sigma$=1.89).

\begin{figure*}
\begin{center}
        
        \resizebox{0.497\hsize}{!}{\includegraphics{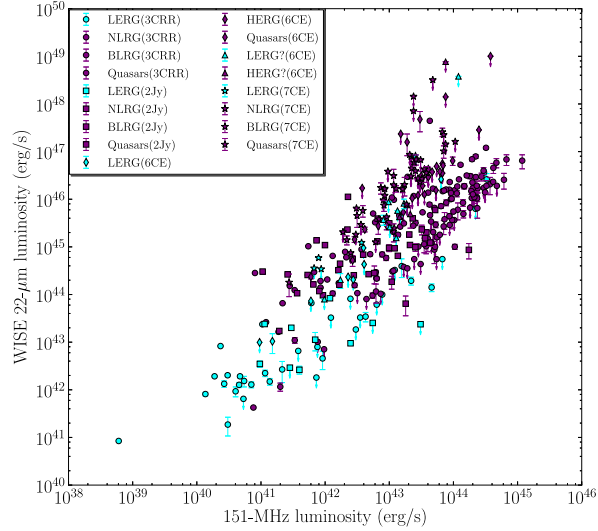}} 
        \caption{Plot of 22$\mu$m luminosity versus 151-MHz radio luminosity. Upper limits are indicated as arrows. In order to show the distribution of HERGs and LERGs clearly, HERGs are plotted as purple filled circles and LERGs are as cyan filled circles.}
\label{lerg_herg}
\end{center}\vspace{-1em}
\end{figure*}

In order to highlight the separate positions of HERGs and LERGs in the radio$-$mid-IR plane, we re-plotted the 22-$\mu$m-, 22-$\mu$m$-$151-MHz luminosities using a different colour scheme. This can be seen in Figure \ref{lerg_herg} where HERGs are plotted with purple colours and LERGs with cyan colours. In this plot, an approximate empirical cutoff between LERGs and HERGs stands out, corresponding to a 22-$\mu$m luminosity around 5$\times$10$^{43}$ erg s$^{-1}$  [$L_{\rm bol} \approx 10^{45}$ erg s$^{-1}$, assuming that the bolometric correction is $\sim$ 20 \citep{2012ref106}]. A similar dividing luminosity was also found by \cite{2006ref20} using a much smaller sample. The objects lying below this line are almost exclusively LERGs. Some overlap between LERGs and NLRGs appears in this plot. Classifications of some sources (especially in the 6CE sample, those labelled with question mark) are not clear which may cause this overlap. Furthermore, as argued by e.g. Mingo et. al. (2013) some low-power objects classified as NLRGs should actually be classified as LERGs. However, it is important to note that almost all LERGs above this cutoff are upper limits. An object that is detected and has a luminosity higher than 5$\times$10$^{43}$ erg s$^{-1}$ is almost certainly a HERG.

\begin{figure*}
\begin{center}
        \resizebox{0.6\hsize}{!}{\includegraphics{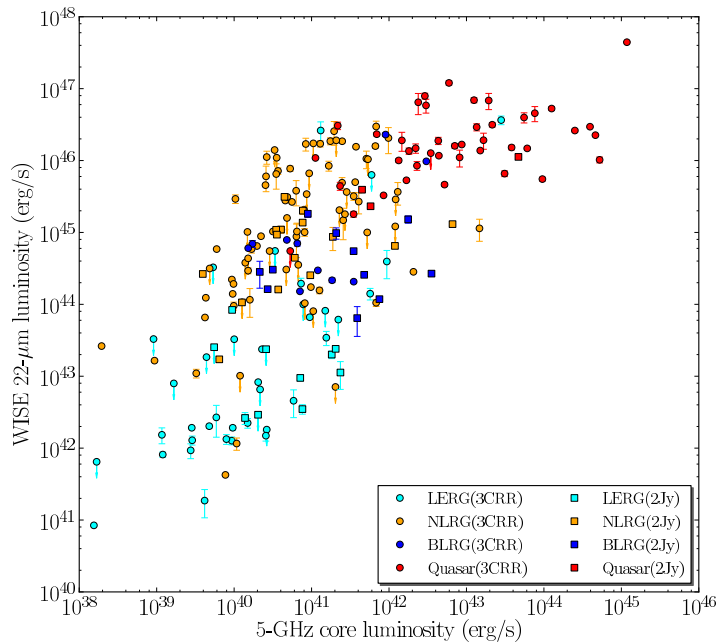}}
        \caption{Plot of 22$\mu$m luminosity versus 5-GHz core luminosity. The 6CE and 7CE samples do not have radio core observations so we only present the 3CRR and 2Jy samples.}
\label{core_ir}
\end{center}\vspace{-1em}
\end{figure*}

The partial correlation test has also been carried out for all objects and each class in the 3CRR and 2Jy samples using the 5-GHz luminosity$-$22-$\mu$m luminosity (Figure \ref{core_ir}) to see if there is any significant correlation between these luminosities. Since only the 3CRR and 2Jy samples have radio core observations, the 6CE and 7CE samples were not used for the analysis. The results of this correlation analysis are also shown in Table \ref{partial_corr}. In Figure \ref{core_ir} NLRGs and BLRGs almost overlap, but quasars have higher luminosities at both wavelengths which can be attributed to some other (non-thermal or emission from the disk) contamination seen in the mid-IR for BLRGs and quasars. Despite the correlation seen for all sources, owing to the effect of redshift dependency, we do not see a significant correlation for each population apart from the LERGs. Our results support the findings of the similar analysis of \cite{2009ref51}. The radio-core luminosity of AGN is an indicator of instantaneous jet power \citep[e.g.][]{1979ref81}. The significant correlation between radio-core emission and mid-IR emission seen in LERGs strengthens the idea that the mid-IR power in LERGs can originate in jets instead of the standard accretion mechanism \citep[e.g.][]{2002ref26}. 

As seen in Figure \ref{lerg_herg}, LERGs exhibit very weak mid-IR emission and some have only upper limits. This can obviously be attributed to lack of obscuring structures. Our results also reinforce the prediction that different accretion modes govern HERGs and LERGs. What drives these different modes is a crucial question. 

\begin{figure*}
\begin{center}
        \resizebox{0.497\hsize}{!}{\includegraphics{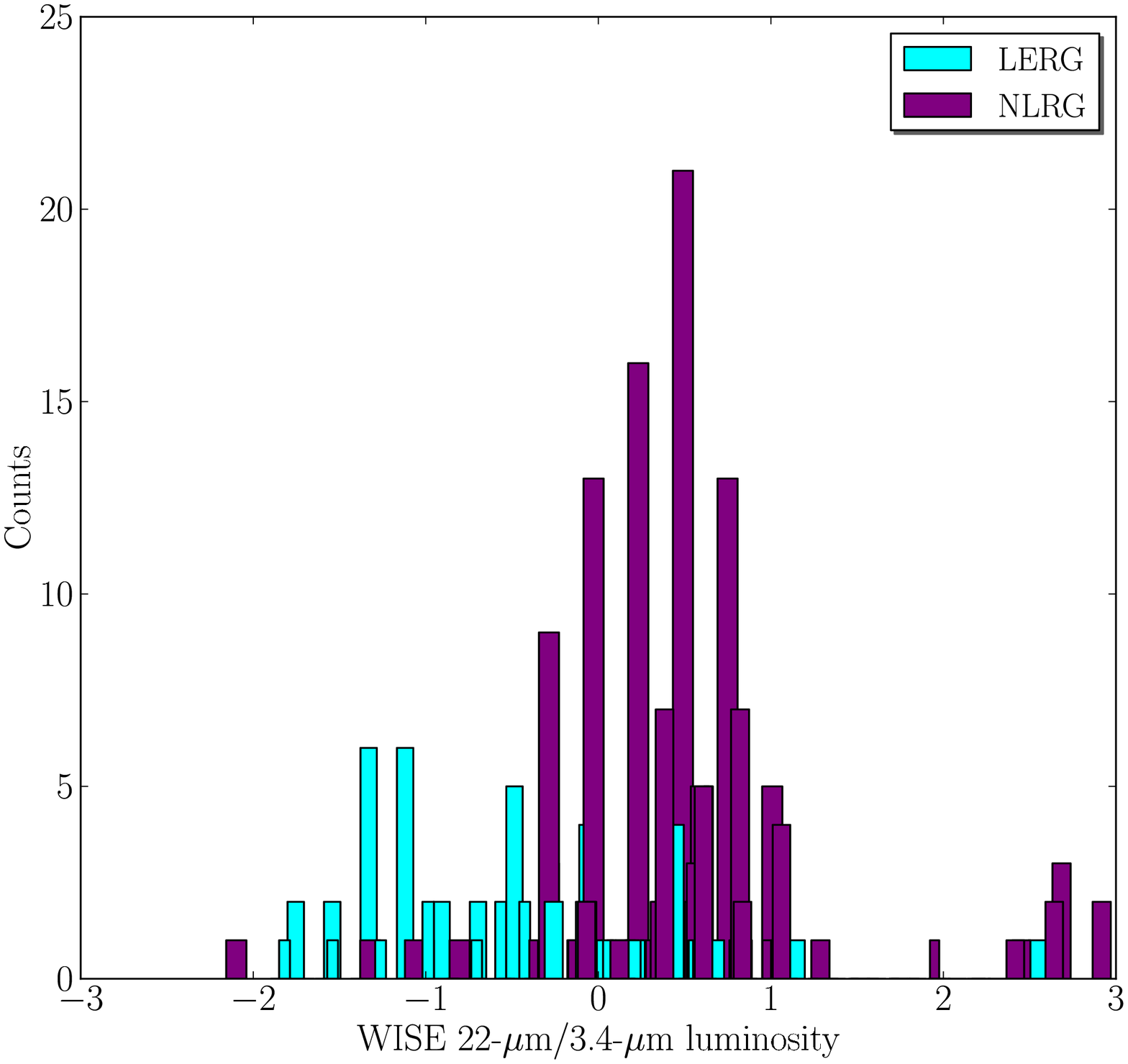}}
        \resizebox{0.497\hsize}{!}{\includegraphics{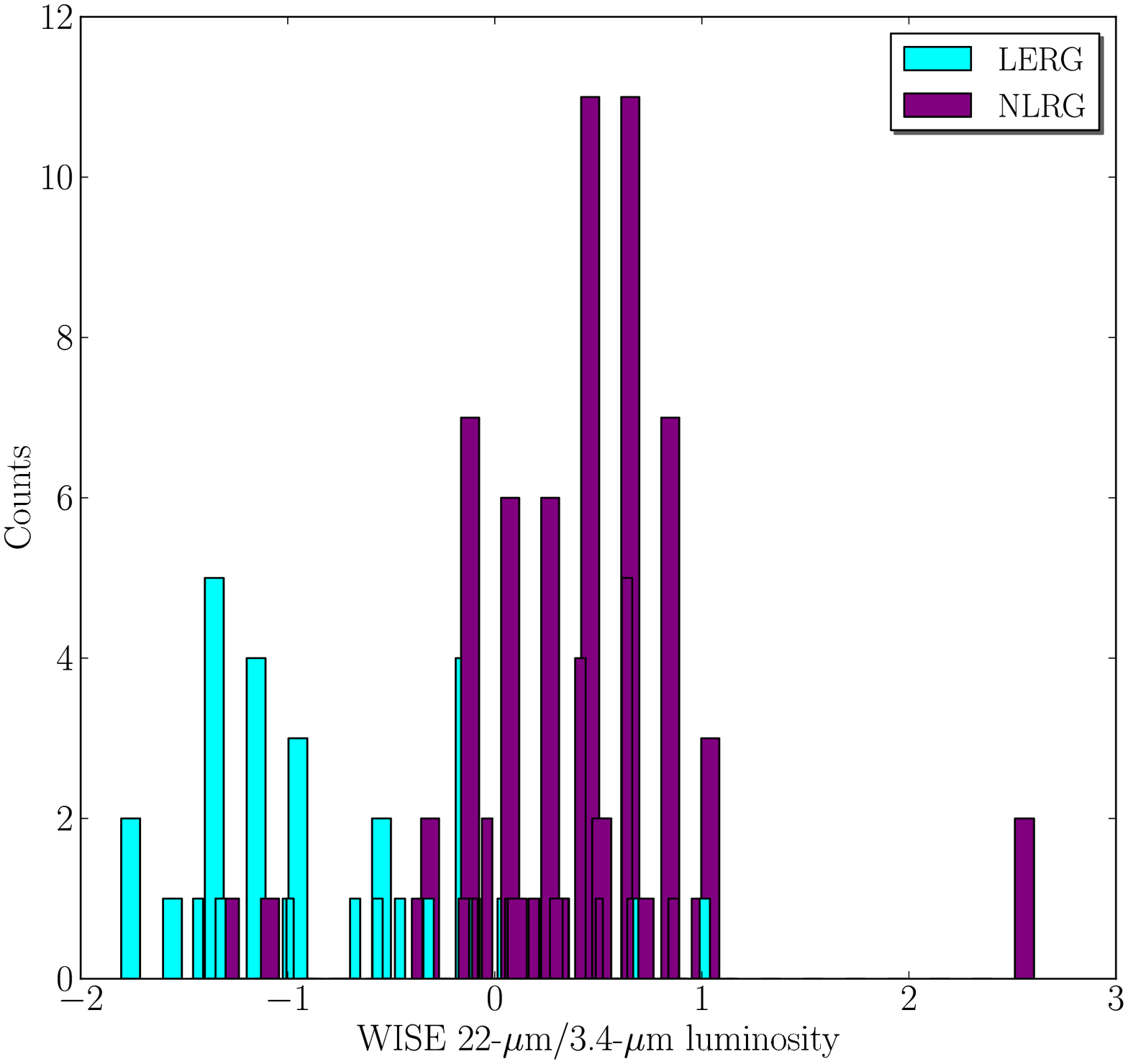}}
        \caption{Histogram of the distribution of 22$\mu$m luminosity over 3.4$\mu$m luminosity (a proxy of the Eddington-scaled accretion rate) versus redshift. In the left plot all sources are used and in the right objects with upper limits are excluded.}
\label{eddington_hist}
\end{center}\vspace{-1em}
\end{figure*}

As discussed in Section 1, one of the hypotheses for the difference between HERGs and LERGs is that there is a limiting value of the Eddington-scaled accretion rate above which radiatively efficient accretion takes place. In order to test this hypothesis we use available $\textit{WISE}$ data; using 22-$\mu$m luminosity as a proxy for the total radiative luminosity of AGN and the 3.6-$\mu$m luminosity for stellar luminosity ($\sim$stellar mass), which is correlated with black-hole mass and thus with the Eddington luminosity for LERGs and NLRGs. Quasars and BLRGs can be contaminated by non-thermal emission in both bands. Moreover, the nuclear quasar emission can overwhelm the stellar light measured at 3.4-$\mu$m. For these reasons, quasars and BLRGs are excluded from the plots. The ratio of 22-$\mu$m to 3.6-$\mu$m then should provide a proxy of the Eddington-scaled accretion rate for NLRGs and LERGs. The results of this test can be seen in Figure \ref{eddington_hist}. In Figure \ref{eddington_hist} (left) all NLRGs and LERGs are used, independent of upper limits. In order to see the effect of upper limits, we also show the distribution of the ratio excluding upper limits (Figure \ref{eddington_hist}-right). Although the result does not change, the overlap between NLRGs and LERGs is reduced in the plot. It is clear that while NLRGs show high values of this ratio ($\sim$ 0.3-10) LERGs have much lower values ($\sim$ 0.01-0.3). Similarly, different accretion rates for HERGs and LERGs were obtained in recent studies \citep[Mingo et al. 2013;][]{2012ref29,2013ref90}. It is worth noting that in both histograms there are some overlaps but since we are only using rough estimators for calculation of the accretion rates this is expected. Furthermore, the classifications of some sources are not clear (such as some sources in the 6CE sample) and this can also lead to some overlap. Nevertheless, this suggests that the 22-$\mu$m/3.4-$\mu$m luminosity ratio can give a good empirical NLRG/LERG classification. A more detailed discussion of the Eddington-scaled accretion rate in radio galaxies is given by Mingo et al. (2013) and Fernandes et al. (in preparation).

\begin{table*}
\begin{tabular}{ccccccc}
\hline
Abscissa&Ordinate&Subsample&Number&Correlation&$\tau$/$\sigma$\\
\hline
L$_{151}$&L$_{22\mu m}$&All&335&Yes&7.97\\
&&LERGs&69&No&1.89\\
&&HERGs&266&Yes&6.08\\
L$_{5}$&L$_{22\mu m}$&All&219&Yes&9.38\\
&&LERGs&47&Yes&4.67\\
&&BLRGs&23&No&-0.12\\ 
&&NLRGs&101&No&2.41\\
&&Quasars&48&No&2.96\\
\hline
\end{tabular}
\caption[]{Results of partial correlation analyses. All sources and subsamples that have relevant luminosities used for the analysis are given in column 3. The number of objects included in the analysis can be seen in column 4. $\tau$/$\sigma$ gives an indication of the strength of the partial correlation in the presence of redshift; a cutoff of $\tau$/$\sigma$ is adopted as $\tau$/$\sigma$$>$3 for a significant correlation. }{\label{partial_corr}}
\end{table*}

%%%%%%%%%%%%%%%%%%%%%%%%%%%%%%%%%%%%%%%%%%%%%%
\section{Summary and Conclusions}
We have used $\textit{WISE}$ to establish the near and mid-IR properties of four radio samples; 3CRR, 2Jy, 6CE and 7CE. The main results are as follows.
\itemize

\item{We evaluated our objects in $\textit{WISE}$ colour-colour diagrams. Various $\textit{WISE}$ colour cuts and criteria have been suggested for selecting AGN \citep{2010ref104,2011ref101,2012ref100,2012ref102,2013ref103}. Although these criteria are detecting some AGN, they are not successful for covering all types of AGN. It can be seen in the diagrams (Figure \ref{colour_colour}) that LERGs and many NLRGs cannot be selected using a simple $\textit{WISE}$ colour cut or by other criteria proposed in the literature.}

\item{The near-IR luminosity$-$redshift relation reinforces the picture in which radio galaxies are hosted by giant ellipticals that formed their stars at high redshifts and evolved passively thereafter. At high redshifts (z$>$0.7) 3CRR objects differ from 6CE and 7CE object having higher near-IR luminosities. This suggests that 3CRR galaxies are more massive systems with higher masses of stars at high redshifts compared to 6CE and 7CE radio galaxies.}

 \item{Our investigations of quasar-radio galaxy unification indicate that quasars are systematically more luminous in the mid-IR than radio galaxies, and more so at 12-$\mu$m than 22-$\mu$m. Our results are consistent with the predictions of smooth torus models which show a strong effect of anisotropy \citep[e.g.][]{2005ref97}.

\item{We have shown for the first time with a large complete sample that low- and high- excitation radio-loud AGN have completely different mid-IR luminosities. While LERGs have extremely weak mid-IR luminosities$-$in fact many of them are not detected and have only upper limits$-$HERGs are mostly luminous sources in the mid-IR. Our results obviously favour previously established accretion models \citep[e.g.][]{1979ref22,1994ref96,2006ref28,2006ref20,2007ref27,2012ref68,2012ref29,2013ref90}; LERGs do not hold any conventional AGN properties and accrete in a radiatively inefficient manner, while HERGs are powered by radiatively efficient accretion. This accretion-mode classification can now be $\textit{explicitly}$ identified in the mid-IR$-$radio plane.

The distribution of each population is quite distinct in the 22-$\mu$m$-$151-MHz luminosity plot. An empirical cutoff stands out in the radio-IR plane (Figure \ref{lerg_herg}), which leads to the conclusion that any object below 4-5$\times$10$^{43}$ erg s$^{-1}$ at 22-$\mu$m is a LERG. Classification of radio sources, hitherto, have relied on expensive optical spectroscopy. Here, we propose that $\textit{WISE}$ data can be effectively used to identify radiatively inefficient and efficient radio-loud AGN.} 

\item{One model of the difference between LERGs and HERGs is that there is a limiting value of the Eddington-scaled accretion rate above which radiatively efficient accretion takes place. Using the 22-$\mu$m and 3.4-$\mu$m luminosities we calculated the ratio of 22-$\mu$m/3.4-$\mu$m as a proxy of the Eddington-scaled accretion rates for NLRGs and LERGs. Although there is some overlap, LERGs ($\sim$ 0.01-0.3) and NLRGs ($\sim$ 0.3-10) differ from each other in this ratio.  Different accretion rates for LERGs and HERGs were also found by others \citep[Mingo et al. 2013;][]{2012ref29,2013ref90}. Since the classification of some objects in the 6CE and 7CE sample is not complete and secure, we may expect to have some overlap in the accretion-rate histograms. Other uncertainties such as the effect of different environments on the radio power of radio galaxies as well as the calculation of black hole mass can contribute to this overlap (see Mingo et al. 2013).

\section*{Acknowledgements}
G\"ulay G\"urkan Uygun thanks the University of Hertfordshire for a PhD studentship. This publication makes use of data products from the Wide-field Infrared Survey Explorer, which is a joint project of the University of California, Los Angeles, and the Jet Propulsion Laboratory/California Institute of Technology, funded by the National Aeronautics and Space Administration.

\bibliographystyle{mn2e}
\bibliography{ref}

\begin{thebibliography}{103}
\expandafter\ifx\csname natexlab\endcsname\relax\def\natexlab#1{#1}\fi

\bibitem[{{Akritas} \& {Siebert}(1996)}]{1996ref80}
{Akritas} M.~G., {Siebert} J., 1996, \mnras, 278, 919

\bibitem[{{Allen} {et~al.}(2006){Allen}, {Dunn}, {Fabian}, {Taylor}, \&
  {Reynolds}}]{2006ref33}
{Allen} S.~W., {Dunn} R.~J.~H., {Fabian} A.~C., {Taylor} G.~B., {Reynolds}
  C.~S., 2006, \mnras, 372, 21

\bibitem[{{Allington-Smith} {et~al.}(1982){Allington-Smith}, {Perryman},
  {Longair}, {Gunn}, \& {Westphal}}]{1982ref45}
{Allington-Smith} J.~R., {Perryman} M.~A.~C., {Longair} M.~S., {Gunn} J.~E.,
  {Westphal} J.~A., 1982, \mnras, 201, 331

\bibitem[{{Antonucci}(2012)}]{2012ref120}
{Antonucci} R., 2012, Astronomical and Astrophysical Transactions, 27, 557

\bibitem[{{Antonucci}(1984)}]{1984ref18}
{Antonucci} R.~R.~J., 1984, \apj, 278, 499

\bibitem[{{Antonucci} \& {Miller}(1985)}]{1985ref39}
{Antonucci} R.~R.~J., {Miller} J.~S., 1985, \apj, 297, 621

\bibitem[{{Assef} {et~al.}(2010){Assef}, {Kochanek}, {Brodwin}, {Cool},
  {Forman}, {Gonzalez}, {Hickox}, {Jones}, {Le Floc'h}, {Moustakas}, {Murray},
  \& {Stern}}]{2010ref104}
{Assef} R.~J., {Kochanek} C.~S., {Brodwin} M., {Cool} R., {Forman} W.,
  {Gonzalez} A.~H., {Hickox} R.~C., {Jones} C., {Le Floc'h} E., {Moustakas} J.,
  {Murray} S.~S., {Stern} D., 2010, \apj, 713, 970

\bibitem[{{Assef} {et~al.}(2013){Assef}, {Stern}, {Kochanek}, {Blain},
  {Brodwin}, {Brown}, {Donoso}, {Eisenhardt}, {Jannuzi}, {Jarrett}, {Stanford},
  {Tsai}, {Wu}, \& {Yan}}]{2013ref103}
{Assef} R.~J., {Stern} D., {Kochanek} C.~S., {Blain} A.~W., {Brodwin} M.,
  {Brown} M.~J.~I., {Donoso} E., {Eisenhardt} P.~R.~M., {Jannuzi} B.~T.,
  {Jarrett} T.~H., {Stanford} S.~A., {Tsai} C.-W., {Wu} J., {Yan} L., 2013,
  \apj, 772, 26

\bibitem[{{Baldwin} {et~al.}(1985){Baldwin}, {Boysen}, {Hales}, {Jennings},
  {Waggett}, {Warner}, \& {Wilson}}]{1985ref7}
{Baldwin} J.~E., {Boysen} R.~C., {Hales} S.~E.~G., {Jennings} J.~E., {Waggett}
  P.~C., {Warner} P.~J., {Wilson} D.~M.~A., 1985, \mnras, 217, 717

\bibitem[{{Barthel}(1989)}]{1989ref17}
{Barthel} P.~D., 1989, \apj, 336, 606

\bibitem[{{Bennett}(1962)}]{1962ref1}
{Bennett} A.~S., 1962, \mnras, 125, 75

\bibitem[{{Best} \& {Heckman}(2012)}]{2012ref29}
{Best} P.~N., {Heckman} T.~M., 2012, \mnras, 421, 1569

\bibitem[{{Best} {et~al.}(2006){Best}, {Kaiser}, {Heckman}, \&
  {Kauffmann}}]{2006ref67}
{Best} P.~N., {Kaiser} C.~R., {Heckman} T.~M., {Kauffmann} G., 2006, \mnras,
  368, L67

\bibitem[{{Best} {et~al.}(1998){Best}, {Longair}, \& {Roettgering}}]{1998ref59}
{Best} P.~N., {Longair} M.~S., {Roettgering} H.~J.~A., 1998, \mnras, 295, 549

\bibitem[{{Blandford} \& {K{\"o}nigl}(1979)}]{1979ref81}
{Blandford} R.~D., {K{\"o}nigl} A., 1979, \apj, 232, 34

\bibitem[{{Chiaberge} {et~al.}(2002){Chiaberge}, {Capetti}, \&
  {Celotti}}]{2002ref26}
{Chiaberge} M., {Capetti} A., {Celotti} A., 2002, \aap, 394, 791

\bibitem[{{Cleary} {et~al.}(2007){Cleary}, {Lawrence}, {Marshall}, {Hao}, \&
  {Meier}}]{2007ref43}
{Cleary} K., {Lawrence} C.~R., {Marshall} J.~A., {Hao} L., {Meier} D., 2007,
  \apj, 660, 117

\bibitem[{{Dicken} {et~al.}(2009){Dicken}, {Tadhunter}, {Axon}, {Morganti},
  {Inskip}, {Holt}, {Gonz{\'a}lez Delgado}, \& {Groves}}]{2009ref10}
{Dicken} D., {Tadhunter} C., {Axon} D., {Morganti} R., {Inskip} K.~J., {Holt}
  J., {Gonz{\'a}lez Delgado} R., {Groves} B., 2009, \apj, 694, 268

\bibitem[{{Dicken} {et~al.}(2012){Dicken}, {Tadhunter}, {Axon}, {Morganti},
  {Robinson}, {Kouwenhoven}, {Spoon}, {Kharb}, {Inskip}, {Holt}, {Ramos
  Almeida}, \& {Nesvadba}}]{2012ref9}
{Dicken} D., {Tadhunter} C., {Axon} D., {Morganti} R., {Robinson} A.,
  {Kouwenhoven} M.~B.~N., {Spoon} H., {Kharb} P., {Inskip} K.~J., {Holt} J.,
  {Ramos Almeida} C., {Nesvadba} N.~P.~H., 2012, \apj, 745, 172

\bibitem[{{Dicken} {et~al.}(2008){Dicken}, {Tadhunter}, {Morganti}, {Buchanan},
  {Oosterloo}, \& {Axon}}]{2008ref11}
{Dicken} D., {Tadhunter} C., {Morganti} R., {Buchanan} C., {Oosterloo} T.,
  {Axon} D., 2008, \apj, 678, 712

\bibitem[{{Eales} {et~al.}(1997){Eales}, {Rawlings}, {Law-Green}, {Cotter}, \&
  {Lacy}}]{1997ref6}
{Eales} S., {Rawlings} S., {Law-Green} D., {Cotter} G., {Lacy} M., 1997,
  \mnras, 291, 593

\bibitem[{{Eales}(1985{\natexlab{a}})}]{1985ref47}
{Eales} S.~A., 1985{\natexlab{a}}, \mnras, 217, 167

\bibitem[{{Eales}(1985{\natexlab{b}})}]{1985ref8}
---, 1985{\natexlab{b}}, \mnras, 217, 149

\bibitem[{{Eales} \& {Rawlings}(1993)}]{1993ref48}
{Eales} S.~A., {Rawlings} S., 1993, \apj, 411, 67

\bibitem[{{Elbaz} {et~al.}(2011){Elbaz}, {Dickinson}, {Hwang},
  {D{\'{\i}}az-Santos}, {Magdis}, {Magnelli}, {Le Borgne}, {Galliano},
  {Pannella}, {Chanial}, {Armus}, {Charmandaris}, {Daddi}, {Aussel}, {Popesso},
  {Kartaltepe}, {Altieri}, {Valtchanov}, {Coia}, {Dannerbauer}, {Dasyra},
  {Leiton}, {Mazzarella}, {Alexander}, {Buat}, {Burgarella}, {Chary}, {Gilli},
  {Ivison}, {Juneau}, {Le Floc'h}, {Lutz}, {Morrison}, {Mullaney}, {Murphy},
  {Pope}, {Scott}, {Brodwin}, {Calzetti}, {Cesarsky}, {Charlot}, {Dole},
  {Eisenhardt}, {Ferguson}, {F{\"o}rster Schreiber}, {Frayer}, {Giavalisco},
  {Huynh}, {Koekemoer}, {Papovich}, {Reddy}, {Surace}, {Teplitz}, {Yun}, \&
  {Wilson}}]{2011ref117}
{Elbaz} D., {Dickinson} M., {Hwang} H.~S., {D{\'{\i}}az-Santos} T., {Magdis}
  G., {Magnelli} B., {Le Borgne} D., {Galliano} F., {Pannella} M., {Chanial}
  P., {Armus} L., {Charmandaris} V., {Daddi} E., {Aussel} H., {Popesso} P.,
  {Kartaltepe} J., {Altieri} B., {Valtchanov} I., {Coia} D., {Dannerbauer} H.,
  {Dasyra} K., {Leiton} R., {Mazzarella} J., {Alexander} D.~M., {Buat} V.,
  {Burgarella} D., {Chary} R.-R., {Gilli} R., {Ivison} R.~J., {Juneau} S., {Le
  Floc'h} E., {Lutz} D., {Morrison} G.~E., {Mullaney} J.~R., {Murphy} E.,
  {Pope} A., {Scott} D., {Brodwin} M., {Calzetti} D., {Cesarsky} C., {Charlot}
  S., {Dole} H., {Eisenhardt} P., {Ferguson} H.~C., {F{\"o}rster Schreiber} N.,
  {Frayer} D., {Giavalisco} M., {Huynh} M., {Koekemoer} A.~M., {Papovich} C.,
  {Reddy} N., {Surace} C., {Teplitz} H., {Yun} M.~S., {Wilson} G., 2011, \aap,
  533, A119

\bibitem[{{Eracleous} \& {Halpern}(2001)}]{2001ref41}
{Eracleous} M., {Halpern} J.~P., 2001, \apj, 554, 240

\bibitem[{{Evans} {et~al.}(2006){Evans}, {Worrall}, {Hardcastle}, {Kraft}, \&
  {Birkinshaw}}]{2006ref28}
{Evans} D.~A., {Worrall} D.~M., {Hardcastle} M.~J., {Kraft} R.~P., {Birkinshaw}
  M., 2006, \apj, 642, 96

\bibitem[{{Feigelson} \& {Nelson}(1985)}]{1985ref109}
{Feigelson} E.~D., {Nelson} P.~I., 1985, \apj, 293, 192

\bibitem[{{Fernandes} {et~al.}(2011){Fernandes}, {Jarvis}, {Rawlings},
  {Mart{\'{\i}}nez-Sansigre}, {Hatziminaoglou}, {Lacy}, {Page}, {Stevens}, \&
  {Vardoulaki}}]{2011ref114}
{Fernandes} C.~A.~C., {Jarvis} M.~J., {Rawlings} S., {Mart{\'{\i}}nez-Sansigre}
  A., {Hatziminaoglou} E., {Lacy} M., {Page} M.~J., {Stevens} J.~A.,
  {Vardoulaki} E., 2011, \mnras, 411, 1909

\bibitem[{{Ferrarese} \& {Merritt}(2000)}]{2000ref66}
{Ferrarese} L., {Merritt} D., 2000, \apjl, 539, L9

\bibitem[{{Grimes} {et~al.}(2003){Grimes}, {Rawlings}, \&
  {Willott}}]{2003ref93}
{Grimes} J.~A., {Rawlings} S., {Willott} C.~J., 2003, \nar, 47, 205

\bibitem[{{Haas} {et~al.}(2004){Haas}, {M{\"u}ller}, {Bertoldi}, {Chini},
  {Egner}, {Freudling}, {Klaas}, {Krause}, {Lemke}, {Meisenheimer},
  {Siebenmorgen}, \& {van Bemmel}}]{2004ref70}
{Haas} M., {M{\"u}ller} S.~A.~H., {Bertoldi} F., {Chini} R., {Egner} S.,
  {Freudling} W., {Klaas} U., {Krause} O., {Lemke} D., {Meisenheimer} K.,
  {Siebenmorgen} R., {van Bemmel} I., 2004, \aap, 424, 531

\bibitem[{{Haas} {et~al.}(2005){Haas}, {Siebenmorgen}, {Schulz}, {Kr{\"u}gel},
  \& {Chini}}]{2005ref71}
{Haas} M., {Siebenmorgen} R., {Schulz} B., {Kr{\"u}gel} E., {Chini} R., 2005,
  \aap, 442, L39

\bibitem[{{Haas} {et~al.}(2008){Haas}, {Willner}, {Heymann}, {Ashby}, {Fazio},
  {Wilkes}, {Chini}, \& {Siebenmorgen}}]{2008ref86}
{Haas} M., {Willner} S.~P., {Heymann} F., {Ashby} M.~L.~N., {Fazio} G.~G.,
  {Wilkes} B.~J., {Chini} R., {Siebenmorgen} R., 2008, \apj, 688, 122

\bibitem[{{Hardcastle}(2004)}]{2004ref35}
{Hardcastle} M.~J., 2004, \aap, 414, 927

\bibitem[{{Hardcastle} {et~al.}(2006){Hardcastle}, {Evans}, \&
  {Croston}}]{2006ref25}
{Hardcastle} M.~J., {Evans} D.~A., {Croston} J.~H., 2006, \mnras, 370, 1893

\bibitem[{{Hardcastle} {et~al.}(2007){Hardcastle}, {Evans}, \&
  {Croston}}]{2007ref27}
---, 2007, \mnras, 376, 1849

\bibitem[{{Hardcastle} {et~al.}(2009){Hardcastle}, {Evans}, \&
  {Croston}}]{2009ref51}
---, 2009, \mnras, 396, 1929

\bibitem[{{Hardcastle} \& {Krause}(2013)}]{2013ref82}
{Hardcastle} M.~J., {Krause} M.~G.~H., 2013, \mnras, 430, 174

\bibitem[{{Heckman} {et~al.}(1992){Heckman}, {Chambers}, \&
  {Postman}}]{1992ref69}
{Heckman} T.~M., {Chambers} K.~C., {Postman} M., 1992, \apj, 391, 39

\bibitem[{{Heckman} {et~al.}(1994){Heckman}, {O'Dea}, {Baum}, \&
  {Laurikainen}}]{1994ref56}
{Heckman} T.~M., {O'Dea} C.~P., {Baum} S.~A., {Laurikainen} E., 1994, \apj,
  428, 65

\bibitem[{{Hine} \& {Longair}(1979)}]{1979ref22}
{Hine} R.~G., {Longair} M.~S., 1979, \mnras, 188, 111

\bibitem[{{H{\"o}nig} {et~al.}(2011){H{\"o}nig}, {Leipski}, {Antonucci}, \&
  {Haas}}]{2011ref121}
{H{\"o}nig} S.~F., {Leipski} C., {Antonucci} R., {Haas} M., 2011, \apj, 736, 26

\bibitem[{{Inskip} {et~al.}(2002){Inskip}, {Best}, {Longair}, \&
  {MacKay}}]{2002ref84}
{Inskip} K.~J., {Best} P.~N., {Longair} M.~S., {MacKay} D.~J.~C., 2002, \mnras,
  329, 277

\bibitem[{{Inskip} {et~al.}(2010){Inskip}, {Tadhunter}, {Morganti}, {Holt},
  {Ramos Almeida}, \& {Dicken}}]{2010ref60}
{Inskip} K.~J., {Tadhunter} C.~N., {Morganti} R., {Holt} J., {Ramos Almeida}
  C., {Dicken} D., 2010, \mnras, 407, 1739

\bibitem[{{Isobe} {et~al.}(1986){Isobe}, {Feigelson}, \& {Nelson}}]{1986ref110}
{Isobe} T., {Feigelson} E.~D., {Nelson} P.~I., 1986, \apj, 306, 490

\bibitem[{{Jackson} \& {Rawlings}(1997)}]{1997ref50}
{Jackson} N., {Rawlings} S., 1997, \mnras, 286, 241

\bibitem[{{Janssen} {et~al.}(2012){Janssen}, {R{\"o}ttgering}, {Best}, \&
  {Brinchmann}}]{2012ref68}
{Janssen} R.~M.~J., {R{\"o}ttgering} H.~J.~A., {Best} P.~N., {Brinchmann} J.,
  2012, \aap, 541, A62

\bibitem[{{Jarrett} {et~al.}(2011){Jarrett}, {Cohen}, {Masci}, {Wright},
  {Stern}, {Benford}, {Blain}, {Carey}, {Cutri}, {Eisenhardt}, {Lonsdale},
  {Mainzer}, {Marsh}, {Padgett}, {Petty}, {Ressler}, {Skrutskie}, {Stanford},
  {Surace}, {Tsai}, {Wheelock}, \& {Yan}}]{2011ref101}
{Jarrett} T.~H., {Cohen} M., {Masci} F., {Wright} E., {Stern} D., {Benford} D.,
  {Blain} A., {Carey} S., {Cutri} R.~M., {Eisenhardt} P., {Lonsdale} C.,
  {Mainzer} A., {Marsh} K., {Padgett} D., {Petty} S., {Ressler} M., {Skrutskie}
  M., {Stanford} S., {Surace} J., {Tsai} C.~W., {Wheelock} S., {Yan} D.~L.,
  2011, \apj, 735, 112

\bibitem[{{Jarvis} {et~al.}(2001){Jarvis}, {Rawlings}, {Eales}, {Blundell},
  {Bunker}, {Croft}, {McLure}, \& {Willott}}]{2001ref57}
{Jarvis} M.~J., {Rawlings} S., {Eales} S., {Blundell} K.~M., {Bunker} A.~J.,
  {Croft} S., {McLure} R.~J., {Willott} C.~J., 2001, \mnras, 326, 1585

\bibitem[{{Kormendy} \& {Richstone}(1995)}]{1995ref64}
{Kormendy} J., {Richstone} D., 1995, \araa, 33, 581

\bibitem[{{Kraft} {et~al.}(2007){Kraft}, {Birkinshaw}, {Hardcastle}, {Evans},
  {Croston}, {Worrall}, \& {Murray}}]{2007ref62}
{Kraft} R.~P., {Birkinshaw} M., {Hardcastle} M.~J., {Evans} D.~A., {Croston}
  J.~H., {Worrall} D.~M., {Murray} S.~S., 2007, \apj, 659, 1008

\bibitem[{{Lacy} {et~al.}(2004){Lacy}, {Storrie-Lombardi}, {Sajina},
  {Appleton}, {Armus}, {Chapman}, {Choi}, {Fadda}, {Fang}, {Frayer},
  {Heinrichsen}, {Helou}, {Im}, {Marleau}, {Masci}, {Shupe}, {Soifer},
  {Surace}, {Teplitz}, {Wilson}, \& {Yan}}]{2004ref98}
{Lacy} M., {Storrie-Lombardi} L.~J., {Sajina} A., {Appleton} P.~N., {Armus} L.,
  {Chapman} S.~C., {Choi} P.~I., {Fadda} D., {Fang} F., {Frayer} D.~T.,
  {Heinrichsen} I., {Helou} G., {Im} M., {Marleau} F.~R., {Masci} F., {Shupe}
  D.~L., {Soifer} B.~T., {Surace} J., {Teplitz} H.~I., {Wilson} G., {Yan} L.,
  2004, \apjs, 154, 166

\bibitem[{{Laing} {et~al.}(1994){Laing}, {Jenkins}, {Wall}, \&
  {Unger}}]{1994ref96}
{Laing} R.~A., {Jenkins} C.~R., {Wall} J.~V., {Unger} S.~W., 1994, in
  Astronomical Society of the Pacific Conference Series, Vol.~54, The Physics
  of Active Galaxies, {Bicknell} G.~V., {Dopita} M.~A., {Quinn} P.~J., eds., p.
  201

\bibitem[{{Laing} {et~al.}(1983){Laing}, {Riley}, \& {Longair}}]{1983ref2}
{Laing} R.~A., {Riley} J.~M., {Longair} M.~S., 1983, \mnras, 204, 151

\bibitem[{{Lawrence}(1991)}]{1991ref91}
{Lawrence} A., 1991, \mnras, 252, 586

\bibitem[{{Lilly}(1989)}]{1989ref46}
{Lilly} S.~J., 1989, \apj, 340, 77

\bibitem[{{Lilly} \& {Longair}(1982)}]{1982ref58}
{Lilly} S.~J., {Longair} M.~S., 1982, in IAU Symposium, Vol.~97, Extragalactic
  Radio Sources, {Heeschen} D.~S., {Wade} C.~M., eds., pp. 413--421

\bibitem[{{Madejski} {et~al.}(2000){Madejski}, {{\.Z}ycki}, {Done}, {Valinia},
  {Blanco}, {Rothschild}, \& {Turek}}]{2000ref40}
{Madejski} G., {{\.Z}ycki} P., {Done} C., {Valinia} A., {Blanco} P.,
  {Rothschild} R., {Turek} B., 2000, \apjl, 535, L87

\bibitem[{{Magorrian} {et~al.}(1998){Magorrian}, {Tremaine}, {Richstone},
  {Bender}, {Bower}, {Dressler}, {Faber}, {Gebhardt}, {Green}, {Grillmair},
  {Kormendy}, \& {Lauer}}]{1998ref65}
{Magorrian} J., {Tremaine} S., {Richstone} D., {Bender} R., {Bower} G.,
  {Dressler} A., {Faber} S.~M., {Gebhardt} K., {Green} R., {Grillmair} C.,
  {Kormendy} J., {Lauer} T., 1998, \aj, 115, 2285

\bibitem[{{Mart{\'{\i}}nez-Sansigre} \& {Rawlings}(2011)}]{2011ref32}
{Mart{\'{\i}}nez-Sansigre} A., {Rawlings} S., 2011, \mnras, 414, 1937

\bibitem[{{Mateos} {et~al.}(2012){Mateos}, {Alonso-Herrero}, {Carrera},
  {Blain}, {Watson}, {Barcons}, {Braito}, {Severgnini}, {Donley}, \&
  {Stern}}]{2012ref100}
{Mateos} S., {Alonso-Herrero} A., {Carrera} F.~J., {Blain} A., {Watson} M.~G.,
  {Barcons} X., {Braito} V., {Severgnini} P., {Donley} J.~L., {Stern} D., 2012,
  \mnras, 426, 3271

\bibitem[{{McLure} \& {Dunlop}(2000)}]{2000ref85}
{McLure} R.~J., {Dunlop} J.~S., 2000, \mnras, 317, 249

\bibitem[{{McLure} \& {Dunlop}(2001)}]{2001ref108}
---, 2001, \mnras, 327, 199

\bibitem[{{McLure} {et~al.}(2004){McLure}, {Willott}, {Jarvis}, {Rawlings},
  {Hill}, {Mitchell}, {Dunlop}, \& {Wold}}]{2004ref107}
{McLure} R.~J., {Willott} C.~J., {Jarvis} M.~J., {Rawlings} S., {Hill} G.~J.,
  {Mitchell} E., {Dunlop} J.~S., {Wold} M., 2004, \mnras, 351, 347

\bibitem[{{McNamara} {et~al.}(2011){McNamara}, {Rohanizadegan}, \&
  {Nulsen}}]{2011ref31}
{McNamara} B.~R., {Rohanizadegan} M., {Nulsen} P.~E.~J., 2011, \apj, 727, 39

\bibitem[{{Meisenheimer} {et~al.}(2001){Meisenheimer}, {Haas}, {M{\"u}ller},
  {Chini}, {Klaas}, \& {Lemke}}]{2001ref21}
{Meisenheimer} K., {Haas} M., {M{\"u}ller} S.~A.~H., {Chini} R., {Klaas} U.,
  {Lemke} D., 2001, \aap, 372, 719

\bibitem[{{Merloni} \& {Heinz}(2007)}]{2007ref30}
{Merloni} A., {Heinz} S., 2007, \mnras, 381, 589

\bibitem[{{Miller} \& {Antonucci}(1983)}]{1983ref38}
{Miller} J.~S., {Antonucci} R.~R.~J., 1983, \apjl, 271, L7

\bibitem[{{Narayan} \& {Yi}(1995)}]{1995ref34}
{Narayan} R., {Yi} I., 1995, \apj, 452, 710

\bibitem[{{Nenkova} {et~al.}(2002){Nenkova}, {Ivezi{\'c}}, \&
  {Elitzur}}]{2002ref77}
{Nenkova} M., {Ivezi{\'c}} {\v Z}., {Elitzur} M., 2002, \apjl, 570, L9

\bibitem[{{Nenkova} {et~al.}(2008{\natexlab{a}}){Nenkova}, {Sirocky},
  {Ivezi{\'c}}, \& {Elitzur}}]{2008ref78}
{Nenkova} M., {Sirocky} M.~M., {Ivezi{\'c}} {\v Z}., {Elitzur} M.,
  2008{\natexlab{a}}, \apj, 685, 147

\bibitem[{{Nenkova} {et~al.}(2008{\natexlab{b}}){Nenkova}, {Sirocky},
  {Nikutta}, {Ivezi{\'c}}, \& {Elitzur}}]{2008ref79}
{Nenkova} M., {Sirocky} M.~M., {Nikutta} R., {Ivezi{\'c}} {\v Z}., {Elitzur}
  M., 2008{\natexlab{b}}, \apj, 685, 160

\bibitem[{{Ogle} {et~al.}(2006){Ogle}, {Whysong}, \& {Antonucci}}]{2006ref20}
{Ogle} P., {Whysong} D., {Antonucci} R., 2006, \apj, 647, 161

\bibitem[{{Ogle} {et~al.}(1997){Ogle}, {Cohen}, {Miller}, {Tran}, {Fosbury}, \&
  {Goodrich}}]{1997ref74}
{Ogle} P.~M., {Cohen} M.~H., {Miller} J.~S., {Tran} H.~D., {Fosbury} R.~A.~E.,
  {Goodrich} R.~W., 1997, \apjl, 482, L37

\bibitem[{{Pier} \& {Krolik}(1992)}]{1992ref75}
{Pier} E.~A., {Krolik} J.~H., 1992, \apj, 401, 99

\bibitem[{{Pier} \& {Krolik}(1993)}]{1993ref76}
---, 1993, \apj, 418, 673

\bibitem[{{Punsly} \& {Zhang}(2011)}]{2011ref95}
{Punsly} B., {Zhang} S., 2011, \apjl, 735, L3

\bibitem[{{Rawlings} {et~al.}(2001){Rawlings}, {Eales}, \& {Lacy}}]{2001ref5}
{Rawlings} S., {Eales} S., {Lacy} M., 2001, \mnras, 322, 523

\bibitem[{{Runnoe} {et~al.}(2012){Runnoe}, {Brotherton}, \&
  {Shang}}]{2012ref106}
{Runnoe} J.~C., {Brotherton} M.~S., {Shang} Z., 2012, \mnras, 426, 2677

\bibitem[{{Russell} {et~al.}(2013){Russell}, {McNamara}, {Edge}, {Hogan},
  {Main}, \& {Vantyghem}}]{2013ref90}
{Russell} H.~R., {McNamara} B.~R., {Edge} A.~C., {Hogan} M.~T., {Main} R.~A.,
  {Vantyghem} A.~N., 2013, \mnras

\bibitem[{{Schartmann} {et~al.}(2005){Schartmann}, {Meisenheimer}, {Camenzind},
  {Wolf}, \& {Henning}}]{2005ref97}
{Schartmann} M., {Meisenheimer} K., {Camenzind} M., {Wolf} S., {Henning} T.,
  2005, \aap, 437, 861

\bibitem[{{Schmitt}(1985)}]{1985ref111}
{Schmitt} J.~H.~M.~M., 1985, \apj, 293, 178

\bibitem[{{Shi} {et~al.}(2005){Shi}, {Rieke}, {Hines}, {Neugebauer},
  {Blaylock}, {Rigby}, {Egami}, {Gordon}, \& {Alonso-Herrero}}]{2005ref52}
{Shi} Y., {Rieke} G.~H., {Hines} D.~C., {Neugebauer} G., {Blaylock} M., {Rigby}
  J., {Egami} E., {Gordon} K.~D., {Alonso-Herrero} A., 2005, \apj, 629, 88

\bibitem[{{Siebenmorgen} {et~al.}(2005){Siebenmorgen}, {Haas}, {Kr{\"u}gel}, \&
  {Schulz}}]{2005ref87}
{Siebenmorgen} R., {Haas} M., {Kr{\"u}gel} E., {Schulz} B., 2005, \aap, 436, L5

\bibitem[{{Simpson}(1998)}]{1998ref92}
{Simpson} C., 1998, \mnras, 297, L39

\bibitem[{{Stern} {et~al.}(2012){Stern}, {Assef}, {Benford}, {Blain}, {Cutri},
  {Dey}, {Eisenhardt}, {Griffith}, {Jarrett}, {Lake}, {Masci}, {Petty},
  {Stanford}, {Tsai}, {Wright}, {Yan}, {Harrison}, \& {Madsen}}]{2012ref102}
{Stern} D., {Assef} R.~J., {Benford} D.~J., {Blain} A., {Cutri} R., {Dey} A.,
  {Eisenhardt} P., {Griffith} R.~L., {Jarrett} T.~H., {Lake} S., {Masci} F.,
  {Petty} S., {Stanford} S.~A., {Tsai} C.-W., {Wright} E.~L., {Yan} L.,
  {Harrison} F., {Madsen} K., 2012, \apj, 753, 30

\bibitem[{{Stern} {et~al.}(2005){Stern}, {Eisenhardt}, {Gorjian}, {Kochanek},
  {Caldwell}, {Eisenstein}, {Brodwin}, {Brown}, {Cool}, {Dey}, {Green},
  {Jannuzi}, {Murray}, {Pahre}, \& {Willner}}]{2005ref99}
{Stern} D., {Eisenhardt} P., {Gorjian} V., {Kochanek} C.~S., {Caldwell} N.,
  {Eisenstein} D., {Brodwin} M., {Brown} M.~J.~I., {Cool} R., {Dey} A., {Green}
  P., {Jannuzi} B.~T., {Murray} S.~S., {Pahre} M.~A., {Willner} S.~P., 2005,
  \apj, 631, 163

\bibitem[{{Tadhunter} {et~al.}(1993){Tadhunter}, {Morganti}, {di
  Serego-Alighieri}, {Fosbury}, \& {Danziger}}]{1993ref4}
{Tadhunter} C.~N., {Morganti} R., {di Serego-Alighieri} S., {Fosbury} R.~A.~E.,
  {Danziger} I.~J., 1993, \mnras, 263, 999

\bibitem[{{Takeuchi} {et~al.}(2005){Takeuchi}, {Buat}, {Iglesias-P{\'a}ramo},
  {Boselli}, \& {Burgarella}}]{2005ref118}
{Takeuchi} T.~T., {Buat} V., {Iglesias-P{\'a}ramo} J., {Boselli} A.,
  {Burgarella} D., 2005, \aap, 432, 423

\bibitem[{{Tasse} {et~al.}(2008){Tasse}, {Best}, {R{\"o}ttgering}, \& {Le
  Borgne}}]{2008ref42}
{Tasse} C., {Best} P.~N., {R{\"o}ttgering} H., {Le Borgne} D., 2008, \aap, 490,
  893

\bibitem[{{Urry} \& {Padovani}(1995)}]{1995ref16}
{Urry} C.~M., {Padovani} P., 1995, \pasp, 107, 803

\bibitem[{{van Breugel} {et~al.}(1999){van Breugel}, {De Breuck}, {Stanford},
  {Stern}, {R{\"o}ttgering}, \& {Miley}}]{1999ref83}
{van Breugel} W., {De Breuck} C., {Stanford} S.~A., {Stern} D.,
  {R{\"o}ttgering} H., {Miley} G., 1999, \apjl, 518, L61

\bibitem[{{Visser} {et~al.}(1995){Visser}, {Riley}, {Roettgering}, \&
  {Waldram}}]{1995ref12}
{Visser} A.~E., {Riley} J.~M., {Roettgering} H.~J.~A., {Waldram} E.~M., 1995,
  \aaps, 110, 419

\bibitem[{{Wall} \& {Peacock}(1985)}]{1985ref3}
{Wall} J.~V., {Peacock} J.~A., 1985, \mnras, 216, 173

\bibitem[{{Weedman} {et~al.}(2006){Weedman}, {Polletta}, {Lonsdale}, {Wilkes},
  {Siana}, {Houck}, {Surace}, {Shupe}, {Farrah}, \& {Smith}}]{2006ref116}
{Weedman} D., {Polletta} M., {Lonsdale} C.~J., {Wilkes} B.~J., {Siana} B.,
  {Houck} J.~R., {Surace} J., {Shupe} D., {Farrah} D., {Smith} H.~E., 2006,
  \apj, 653, 101

\bibitem[{{Werner} {et~al.}(2004){Werner}, {Roellig}, {Low}, {Rieke}, {Rieke},
  {Hoffmann}, {Young}, {Houck}, {Brandl}, {Fazio}, {Hora}, {Gehrz}, {Helou},
  {Soifer}, {Stauffer}, {Keene}, {Eisenhardt}, {Gallagher}, {Gautier}, {Irace},
  {Lawrence}, {Simmons}, {Van Cleve}, {Jura}, {Wright}, \&
  {Cruikshank}}]{2004ref73}
{Werner} M.~W., {Roellig} T.~L., {Low} F.~J., {Rieke} G.~H., {Rieke} M.,
  {Hoffmann} W.~F., {Young} E., {Houck} J.~R., {Brandl} B., {Fazio} G.~G.,
  {Hora} J.~L., {Gehrz} R.~D., {Helou} G., {Soifer} B.~T., {Stauffer} J.,
  {Keene} J., {Eisenhardt} P., {Gallagher} D., {Gautier} T.~N., {Irace} W.,
  {Lawrence} C.~R., {Simmons} L., {Van Cleve} J.~E., {Jura} M., {Wright} E.~L.,
  {Cruikshank} D.~P., 2004, \apjs, 154, 1

\bibitem[{{Whysong} \& {Antonucci}(2004)}]{2004ref19}
{Whysong} D., {Antonucci} R., 2004, \apj, 602, 116

\bibitem[{{Willott} {et~al.}(1998){Willott}, {Rawlings}, {Blundell}, \&
  {Lacy}}]{1998ref13}
{Willott} C.~J., {Rawlings} S., {Blundell} K.~M., {Lacy} M., 1998, \mnras, 300,
  625

\bibitem[{{Willott} {et~al.}(2002){Willott}, {Rawlings}, {Blundell}, {Lacy},
  {Hill}, \& {Scott}}]{2002ref14}
{Willott} C.~J., {Rawlings} S., {Blundell} K.~M., {Lacy} M., {Hill} G.~J.,
  {Scott} S.~E., 2002, \mnras, 335, 1120

\bibitem[{{Willott} {et~al.}(2003){Willott}, {Rawlings}, {Jarvis}, \&
  {Blundell}}]{2003ref15}
{Willott} C.~J., {Rawlings} S., {Jarvis} M.~J., {Blundell} K.~M., 2003, \mnras,
  339, 173

\bibitem[{{Wright} {et~al.}(2010){Wright}, {Eisenhardt}, {Mainzer}, {Ressler},
  {Cutri}, {Jarrett}, {Kirkpatrick}, {Padgett}, {McMillan}, {Skrutskie},
  {Stanford}, {Cohen}, {Walker}, {Mather}, {Leisawitz}, {Gautier}, {McLean},
  {Benford}, {Lonsdale}, {Blain}, {Mendez}, {Irace}, {Duval}, {Liu}, {Royer},
  {Heinrichsen}, {Howard}, {Shannon}, {Kendall}, {Walsh}, {Larsen}, {Cardon},
  {Schick}, {Schwalm}, {Abid}, {Fabinsky}, {Naes}, \& {Tsai}}]{2010ref44}
{Wright} E.~L., {Eisenhardt} P.~R.~M., {Mainzer} A.~K., {Ressler} M.~E.,
  {Cutri} R.~M., {Jarrett} T., {Kirkpatrick} J.~D., {Padgett} D., {McMillan}
  R.~S., {Skrutskie} M., {Stanford} S.~A., {Cohen} M., {Walker} R.~G., {Mather}
  J.~C., {Leisawitz} D., {Gautier} III T.~N., {McLean} I., {Benford} D.,
  {Lonsdale} C.~J., {Blain} A., {Mendez} B., {Irace} W.~R., {Duval} V., {Liu}
  F., {Royer} D., {Heinrichsen} I., {Howard} J., {Shannon} M., {Kendall} M.,
  {Walsh} A.~L., {Larsen} M., {Cardon} J.~G., {Schick} S., {Schwalm} M., {Abid}
  M., {Fabinsky} B., {Naes} L., {Tsai} C.-W., 2010, \aj, 140, 1868

\bibitem[{{Wu} {et~al.}(2012){Wu}, {Hao}, {Jia}, {Zhang}, \&
  {Peng}}]{2012ref115}
{Wu} X.-B., {Hao} G., {Jia} Z., {Zhang} Y., {Peng} N., 2012, \aj, 144, 49

\end{thebibliography}

\label{lastpage}

\end{document}